\documentclass[lettersize,journal]{IEEEtran}
\usepackage{amsmath,amsfonts}
\usepackage{algorithm}
\usepackage{algpseudocode}
\usepackage{array}
\usepackage{textcomp}
\usepackage{stfloats}
\usepackage{url}
\usepackage{verbatim}
\usepackage{graphicx}
\usepackage{cite}
\usepackage{booktabs}
\usepackage{tikz}
\usepackage{pgf-pie} 
\usepackage{subcaption}
\usepackage[
    all=normal,
    floats=tight,
    leading=tight,
    leadingfraction=0.870,
    paragraphs=normal,
    charwidths=normal, 
    mathdisplays=normal,
    tracking=normal,
    lists=normal,
    bibnotes=normal,
    wordspacing=normal
]{savetrees}

\begin{document}

\title{Efficient Hardware Acceleration of Sparsely Active Convolutional Spiking Neural Networks}

\author{Jan~Sommer, M. Akif~\"Ozkan, Oliver~Keszocze~\IEEEmembership{Member,~IEEE}, J\"urgen~Teich,~\IEEEmembership{Fellow,~IEEE}
\thanks{The authors are with the Friedrich-Alexander-Universit\"at
Erlangen-N\"urnberg, 91054 Erlangen, Germany. E-mail: \{jan.sommer, akif.oezkan, oliver.keszoecze, juergen.teich\}@fau.de.}
}


\maketitle
\begin{abstract}
Spiking Neural Networks (SNNs) compute in an event-based matter to achieve a more efficient computation than standard Neural Networks. In SNNs, neuronal outputs (i.e. activations) are not encoded with real-valued activations but with sequences of binary spikes. The motivation of using SNNs over conventional neural networks is rooted in the special computational aspects of spike based processing, especially the very high degree of sparsity of neural output activations. Well established architectures for conventional Convolutional Neural Networks (CNNs) feature large spatial arrays of Processing Elements (PEs) that remain highly underutilized in the face of activation sparsity. We propose a novel architecture that is optimized for the processing of Convolutional SNNs (CSNNs) that feature a high degree of activation sparsity. In our architecture, the main strategy is to use less but highly utilized PEs. The PE array used to perform the convolution is only as large as the kernel size, allowing all PEs to be active as long as there are spikes to process. This constant flow of spikes is ensured by compressing the feature maps (i.e. the activations) into queues that can then be processed spike by spike. This compression is performed in run-time using dedicated circuitry, leading to a self-timed scheduling. This allows the processing time to scale directly with the number of spikes. A novel memory organization scheme called \emph{memory interlacing} is used to efficiently store and retrieve the membrane potentials of the individual neurons using multiple small parallel on-chip RAMs. Each RAM is hardwired to its PE, reducing switching circuitry and allowing RAMs to be located in close proximity to the respective PE. We implemented the proposed architecture on an FPGA and achieved a significant speedup compared to other implementations while needing less hardware resources and maintaining a lower energy consumption. 
\end{abstract}

\begin{IEEEkeywords}
Spiking Convolutional Neural Networks (SNN), Hardware Acceleration, Event-Based Processing, Field-Programmable Gate Array (FPGA).
\end{IEEEkeywords}

\section{Introduction}
Artificial Neural Networks (ANNs) have become the go-to solution for many machine learning problems~\cite{SCHMIDHUBER.2015, LeCun.2015}. Generally, ANNs start to outperform conventional machine learning approaches when large amounts of training data are available~\cite{LeCun.2015}. However, this performance comes at a significant computational cost. For example, the ANN model \mbox{ResNet-50} requires a total $3.9 \cdot 10^9$ operations to process a single $224 \times 224$ image~\cite{Sze.2017}. Generally, a trend can be observed that ANNs increase in size as their classification accuracy improves and the task at hand gets more complex~\cite{Sze.2017}. This places a heavy load on the underlying compute resources in terms on memory, memory bandwidth and processing power. To satisfy non-functional requirements such as power, throughput and latency, careful co-design of the underlying algorithms and the respective processing hardware is necessary~\cite{Cambricon.2018, Sze.2017b}. To find more efficient processing systems, inspiration may come from the most efficient cognitive system known to mankind: the human brain. An emerging trend is to implement dedicated hardware to process biologically inspired Spiking Neural Networks (SNNs)~\cite{Bouvier.2019, Mostafa.2017, Fang.2020, Davies.2018, Guo.2019, TrueNorth.2015, SpinalFlow.2020}. The term SNN refers to a large set of models that share one property: the outputs of the neurons (activations) are not encoded with real-valued scalars like in standard NNs, but with sequences of binary events called spikes. What makes SNNs interesting from a computational perspective is their inherent event-driven processing: computations need to be performed only when spikes, i.e., events, occur~\cite{Yousefzadeh.2019, Rueckauer.272018}. To actually achieve a performance advantage compared to standard NNs, three aspects are crucial: 
\begin{itemize}
    \item The neural code determines how information is encoded with binary spikes. The length of the encoding window and the number of spikes required to encode neuronal activations are the most important determinants of the SNN's inference speed and efficiency~\cite{Thorpe.2001, Guo.2021}. In general, the higher the spike sparsity, the better.
    \item In general, high sparsity in the neuronal output activations requires less computations that need to be performed during inference. While spike sparsity is a nice theoretical property, it is actually very difficult to exploit with standard computer architectures, due to the irregular dataflow associated with it~\cite{Shin.2021}.
    \item The output of a spiking neuron does not depend only on its input but also on its internal state called membrane potential. The real-valued membrane potentials need to be stored and thus increase memory requirements which are already very high to begin with. Strategies have to be deployed to multiplex the membrane potential memory to decrease the overall memory footprint. 
\end{itemize}

\IEEEpubidadjcol 
To achieve state-of-the-art classification performance on computer vision tasks, established methods from standard NNs have to be adapted to SNNs. These methods are primarily: convolutional layers and pooling layers. To accelerate such Convolutional Spiking Neural Networks (CSNNs) using specialized hardware, most authors propose large spatial arrays of Processing Elements (PEs)~\cite{Wang.2020, Kang.2020, Davies.2018}. Spatial architectures couple PEs in such away that they can exchange intermediate results without having to access a central memory~\cite{Sze.2017b}. Typical implementations use either fixed data path connections between the individual PEs (Systolic Arrays)~\cite{Guo.2019} or Network-on-Chips (NoCs) that feature a highly flexible packet-based interconnect~\cite{Davies.2018, TrueNorth.2015}. 
Systolic arrays are excellent for performing convolutions in cases where dataflow is easily predictable, i.e., in low sparsity situations~\cite{Chen.EyerissV2}. NoCs are better at handling unpredictable dataflows since they allow balancing the workload over the different PEs. This comes at the cost of the more expensive NoC communication infrastructure for implementing routers and control circuitry. 
The major downside of such spatial architectures is that most PEs are left idle if highly sparse activations have to be processed. However, idle PEs still consume power due to leakage and clock switching (the latter only applies for synchronous implementations). Furthermore, idle PEs are not contributing to the end result and are thus wasting chip area. This is complicated by the fact that activation sparsity and the associated irregular dataflow cannot be predicted a priori~\cite{Cambricon.2018, SCNN.2017}. Consequently, the non-trivial task of mapping neural operations to PEs must be performed at run-time. 
The main contributions of this paper are as follows:
\begin{itemize}
    \item A non-spatial hardware architecture optimized for sparse event-based spike processing using a highly efficient neural code. 
    \item A sequential processing scheme that allows the memory for storing the membrane potentials to be multiplexed, keeping the memory footprint low.
    \item The introduction of a novel memory mapping scheme called \emph{memory interlacing} that allows a highly parallel, fine grained and high-bandwidth distribution of on-chip RAM.
    \item A queue-based self-timed processing to enable a maximal utilization of PEs, because skipping zero-activations is inherent to the dataflow of the architecture. This allows the execution time to scale directly with the number of spikes. The core idea is to employ less PEs but to maximize their utilization.
\end{itemize} 

\section{Background on Spiking Neural Networks}\label{sec:Background}
The large variety of SNN-models has arisen due to trade-offs between biological plausibility and model complexity. In this work, we use the integrate-and-fire (IF)-model~\cite{Izhikevich.2004}. The (IF)-model has the least neuro-computational features of real neurons but is very efficient in its implementation~\cite{Izhikevich.2004}. Recent advances have shown the classification performance of SNNs deploying the simple IF-model to be on par with state of the art non-spiking NNs implementations~\cite{Sengupta.2019, Rueckauer.2017}. For example, Sengupta et al. report an error increment of only 0.15\% on the CIFAR-10 dataset and an error increment of 0.38\% on the difficult ImageNet dataset when using an SNN over a standard NN~\cite{Sengupta.2019}. 

\subsection{The Integrate-and-Fire model}
Mathematically, the \emph{time discrete} IF-model is described as follows: A binary spike from the previous layer $l-1$ arrives at the synapse $i$ of a neuron $j$ and is weighted with the synaptic weight $w_i$. The weighted spike is then integrated (i.e. added) into the neurons membrane potential $V_{m_{j}}^{l}$.
When a membrane potential $V_m$ exceeds the threshold $V_t$, then the neuron fires a spike itself and $V_m$ is reset to 0. 
The membrane potential of a neuron $j$ at layer $l$ at each time step $t$ is described as:
\begin{equation}\label{eq:mem_pot}
V_{m_{j}}^{l}(t)=\left\{\begin{array}{c}
0 \text { if } V_{m_{j}}^{l}(t-1)>V_{t} \\
V_{m_{j}}^{l}(t-1)+\sum_{i} w_{i, j} \cdot x_{i}^{l-1}(t-1) \text { otherwise }
\end{array}\right.
\end{equation} 
The neuron output $x$ is defined by an all-or-nothing threshold activation function:
\begin{equation}\label{eq:thresholding}
x_{j}^{l}(t)= \begin{cases}
1 \text { if } V_{m_{j}}^{l}(t)>V_{t} \\
0 \text { otherwise }
\end{cases}
\end{equation}

Eqns.~\eqref{eq:thresholding} and \eqref{eq:mem_pot} have the following implications:
\begin{itemize}
	\item SNNs are inherently temporal. Their internal state and thus their output is dependent on the time steps $t$.
	\item SNNs operate in an event driven manner. They update their internal state $V_m$ only when an event (i.e. a spike) is presented to them. 
	\item The all-or-nothing thresholding leads to a high degree of spike \emph{sparsity}. That means that the occurrence of spikes is a rare event.
\end{itemize}

The motivation to use SNNs over standard NNs is rooted in the special computational aspects of spike-based processing:
\begin{itemize}
    \item \textbf{No multiplications}. The spikes that encode neuron activations are binary in nature. Weighting the binary activations $\in \{1,0\}$ does not require an actual multiplication, as the multiplication reduces to: $1 \cdot w = w$ and $0 \cdot w = 0$. Thus, only adders are required to integrate the weighted spikes on the membrane potential. Adders require significantly less chip area and power compared to multipliers~\cite{Rueckauer.2017}.
    \item \textbf{Less compute operations}. The output activations of SNNs are significantly sparser then those of standard ANNs~\cite{Lee.2020}.  This sparsity increases in deeper layers of the SNN. Ultimately, this potentially results in less compute operations and thus faster and more energy efficient inference.
    \item \textbf{Less memory accesses}. Even more important than reducing the number of operations is the reduction of memory accesses~\cite{Horowitz.2014}. Here, SNNs have an advantage since activations are binary and activations are highly sparse, resulting in less data movement.
\end{itemize}
Despite the interesting properties of SNNs, there are also some problematic aspects that need to be considered when building specialized hardware:
\begin{itemize}
    \item \textbf{Storage of membrane potentials}. Apart from weights and neuron activations,
SNNs require an additional data structure: the real-valued membrane potentials of
the individual neurons need to be stored and modified during inference.
    \item \textbf{Irregular dataflow}. Activation sparsity cannot be predicted a priori and thus needs to be handled during run-time.
    \item \textbf{Multiple forward passes}. To perform inference on a single sample (e.g. an input
image), the entire SNN has to run its forward pass multiple times. This is because
it takes multiple time steps for the spikes to propagate through the SNN. How many steps are required depends on (a) the network depth, (b) the neural code and (c) the neuron model.
 
\end{itemize}

\subsection{Information Encoding with Binary Spikes}
Since spikes are identical to the logical value 1, information is not represented by spike size. Instead, the activation strength is encoded by the timing and/or the amount of the spikes. A lot of effort has been put into researching neural encoding schemes (called neural code) since they are one of the major determinants of an SNNs performance~\cite{Panzeri.2010, Guo.2021, Rueckauer.272018, Han.2020, Fang.2020}. \emph{Rate coding} is the most used encoding scheme. Here, the activation strength is encoded by the firing rate of the neuron~\cite{Guo.2021}. A high firing rate represents a high activation and vice versa. A neuron
decodes an incoming sequence of spikes by performing temporal averaging over a time window to estimate the mean firing rate. With a longer time window, a larger sample size is obtained, resulting in a more accurate fire rate estimate. This makes rate coding a very time consuming process since it takes a considerable amount of time until the estimated mean firing rate has settled to an accurate value~\cite{Panzeri.2010, Guo.2021}. 
\emph{Time-To-First-Spike} (TTFS) encoding was initially proposed because rate coding could not explain the fast processing in the visual cortex~\cite{Panzeri.2010}. In TTFS coding, information is encoded in the precise firing times of individual spikes. For this, the time to first spike is measured, i.e. the time that has passed between the arrival of a stimulus and the emission of the spike. A large activation is encoded as an earlier spike transmission and vice versa. TTFS coding enables fast processing since neuronal activations can be encoded with a single spike~\cite{Rueckauer.2021}. Section~\ref{sec:SNN_design} provides a detailed discussion of the neural coded we implemented.

\subsection{Convolutional SNNs}
For computer vision tasks, well established techniques like pooling and convolutional layers can be adapted from standard NNs to SNNs. The general principle is well known from image processing: a 2D-kernel with dimensions $R_l \times R_l$ is convolved over a, typically much larger, 2D input with dimensions $H_l \times W_l$ for each layer $l$. For simplicity, we consider only quadratic kernels and standard convolutional layers; the basic principles discussed here can be generalized to other kernel shapes or depthwise convolutional layers. The entries of the kernel are the trainable weights. Each output of a neuron (i.e. activation) is a pixel of the resulting 2D output image called a feature map (fmap). A convolutional layer $l$ typically generates a number of $C_l$ output fmaps called channels which can be interpreted as a third dimension. The channels of the input fmap $C_{l-1}$ dictate the number of channels of each kernel $K_l$. The number of kernels $k_l$ determine the number of output channels $C_l$, i.e. $C_l = k_l$. 
For standard spiking convolutional layers, the binary input fmap \mbox{$\mathbf{X}_{l-1}(t) \in \mathbb{B}^{H_{l-1} \times W_{l-1} \times C_{l-1}}$} is convolved with a a number of $k_l$ kernels $K_l \in \mathbb{R}^{R_{l} \times R_{l} \times C_{l-1}}$ to which the bias $b \in \mathbb{R}^{C_{l}}$ is added to get the \emph{update} to the membrane potentials \mbox{$\mathbf{U}_{l}(t) \in \mathbb{R}^{H_{l} \times W_{l} \times C_{l}}$}. The update to the membrane potential $\mathbf{U}_{l}(t)$ is added to the membrane potentials \mbox{$\mathbf{V}_{m,l}(t-1) \in \mathbb{R}^{H_{l} \times W_{l} \times C_{l}}$}. The membrane potential$\mathbf{V}_{m,l}(t)$ is then thresholded with $V_t \in \mathbb{R}$ to get the resulting output fmap $\mathbf{X}_l(t) \in \mathbb{B}^{H_{l} \times W_{l} \times C_{l}}$. This results in the following equations, with $*$ denoting the convolution operation:

\begin{equation}
    \mathbf{V}_{m,l}(t) = \underbrace{\mathbf{X}_{l-1}(t) * K_l + b_l}_{\mathbf{U}_{l}(t)} + \mathbf{V}_{m,l}(t-1)
    \label{eq:snn_conv}
\end{equation}
\begin{equation}
\mathbf{X}_{l}(t)=\begin{cases}
1 \text { if } \mathbf{V}_{m,l}(t)>V_{t} \\
0 \text { otherwise }
\end{cases}
\end{equation}

\begin{figure}[!t]
  \centering
  \includegraphics[width=0.5\linewidth]{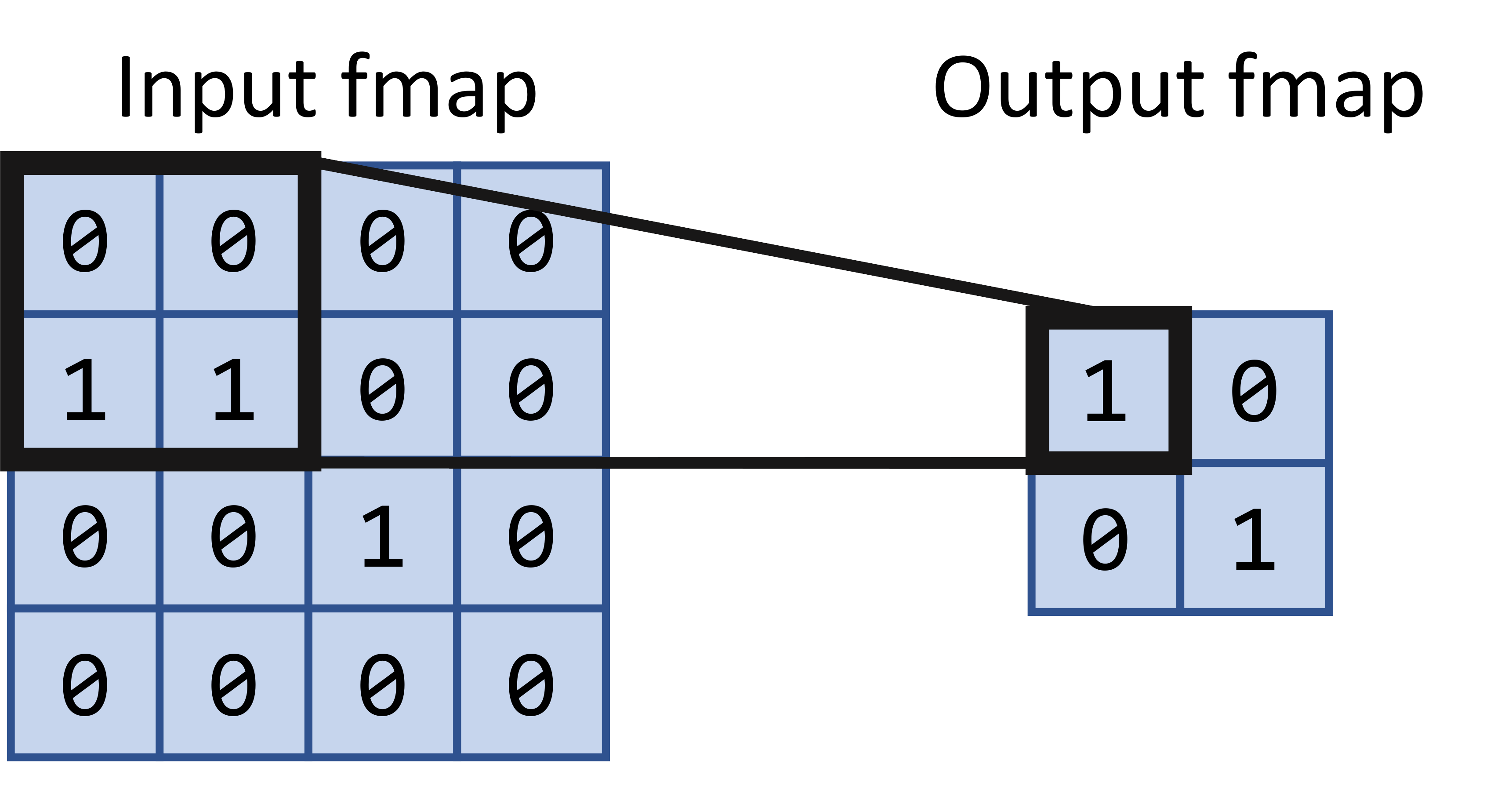}
  \caption{Max-pooling in SNNs, here with a $2 \times 2$ window size.}
  \label{fig:max_pool}
\end{figure}

Pooling can be interpreted as a down-sampling technique to reduce the size of the fmaps. For SNNs, max-pooling is the most used pooling function, due to its simplicity. The pooling operation is performed window-wise in a non-overlapping
fashion. Typical window sizes are $2 \times 2$ and $3 \times 3$.
The implementation of max-pooling for binary fmaps is much simpler than for real-valued
fmaps: searching and selecting the maximum in the pooling window is reduced to
combining all elements in the pooling window with an or-gate (see Fig.~\ref{fig:max_pool}). While solutions to perform average pooling have been proposed, they are much more costly to implement and tend to reduce the classification performance of the SNN \cite{Rueckauer.2017}. 

\section{Related Work}
Davies et al.~\cite{Davies.2018} propose the Intel Loihi, an asynchronous Application Specific Integrated Circuit (ASIC). It features a NoC that interconnects 128 PEs called neurocores. Each neurocore has its own memory and processing unit for implementing 1024 spiking neurons as well as an interface to access the NoC. If a neuron emits a spike, the source neuron puts a packet on the NoC which is then routed to the target neuron. The NoC supports only unicast spike communication, i.e. a neuron can only send a spike to a single target neuron. While the communication itself is asynchronous, the neuronal operations are still conducted in a
time discrete (thus synchronous) way. This requires the neurocores to constantly exchange synchronization
massages to ensure that they are all in the same algorithmic time step. The NoC allows maximum flexibility in the SNN’s connectivity architecture, but it
hinders the exploitation of the characteristic dataflow present in pooling- or
convolutional-layers. It also lacks the ability to explicitly multicast spike packets.
The NoC itself is expensive as it requires a network interface for each neurocore
and multiple routers that have to route the packets while preventing collisions and
deadlocks.

Wang et al.~\cite{Wang.2020} propose SIES, an FPGA-based accelerator with a 2D systolic array for efficiently calculating convolutions. The core idea of systolic arrays is to read data from memory once, but reuse it in multiple PEs so that less memory accesses are required. SIES uses this highly parallel 2D systolic array only to calculate the
\emph{update} of the membrane potential ($\mathbf{U}_{l}(t)$ as per Equation \ref{eq:snn_conv}). This increment is then added to the membrane
potential of each neuron in a sequential way, which appears to be a major bottleneck. 
The systolic array architecture does not harvest the high degree of spike sparsity,
leaving PEs idle.

Kang et al.~\cite{Kang.2020} propose ASIE, an asynchronous ASIC-based convolutional SNN accelerator that implements event-based
processing using the Address Event Representation (AER) protocol. While a 2D fmap in SNNs is basically a (0,1)-matrix $\mathbf{M}$, it is represented in AER by a list of all addresses $(i,j)$ for which \mbox{$\mathbf{M}_{i,j} = 1$}. The 2D array of PEs is ideally as large as the fmap to be processed because each PE implements a physical neuron. For each address event, only the number PEs defined by the kernel size are actually utilized, leaving most PEs idle. E.g. a $30 \times 30$ PE array only utilizes 9 PEs for processing a layer with a $3 \times 3$ kernel.

Fang et al.~\cite{Fang.2020} use a technique called High Level Synthesis (HLS) to describe the SNNs dataflow as a network of recursive filters and automatically synthesize it to an FPGA implementation. The convolution operations themselves are performed using a standard, Multiply-Accumulate-based matrix multiplication unit. A deployment of a temporal, non-rate based neural code is deployed to achieve a very energy efficient implementation. 

\section{Spiking Neural network Design} \label{sec:SNN_design}
\begin{figure}[!t] 
  \centering
  \includegraphics[width=0.99\linewidth]{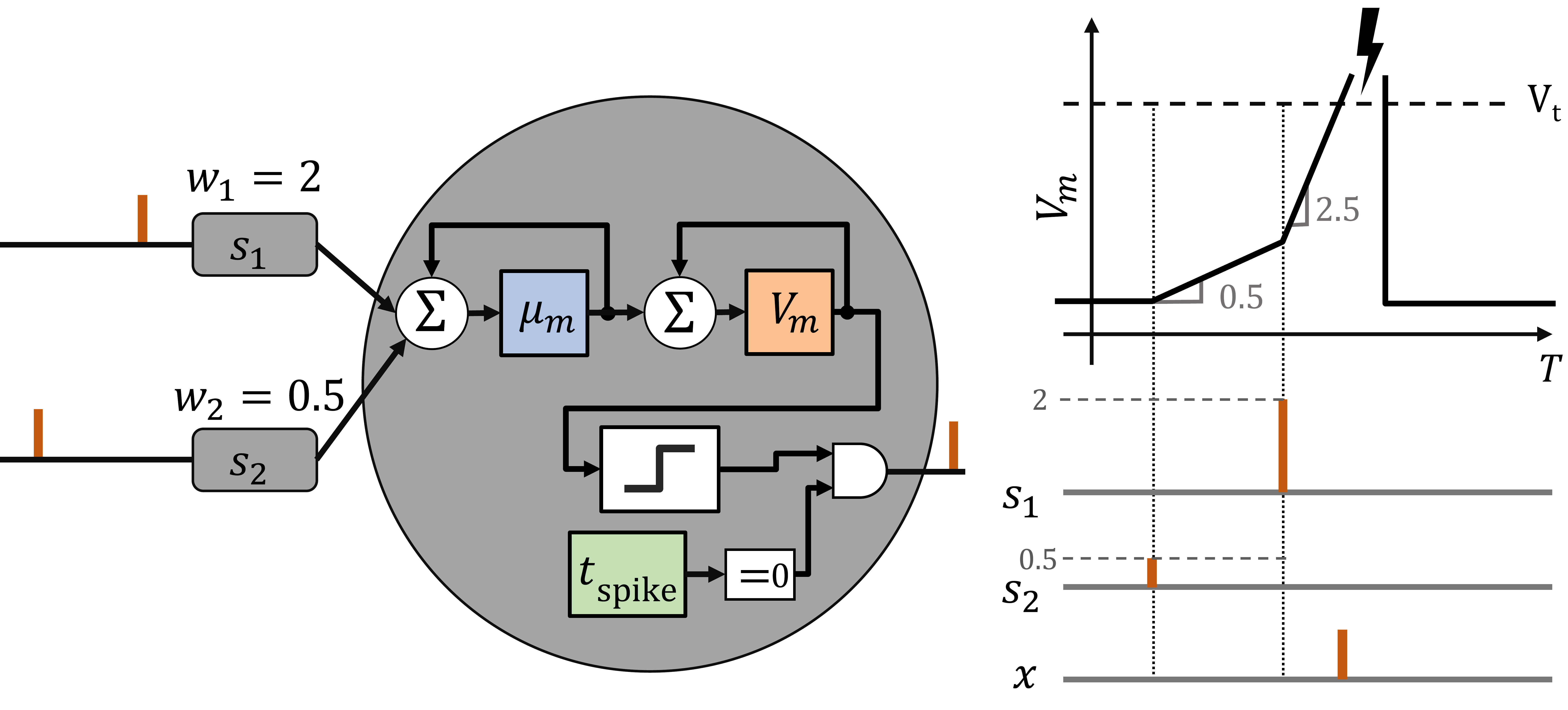}
  \caption{Behavior of a TTFS encoded IF-neuron over time. The neuron integrates the
weighted input spikes on the membrane potential slope $\mu_m$. This slope dictates
the rise or fall of the membrane potential $V_m$ over time. When $V_m$ reaches the
firing threshold $V_t$ and no spike has been emitted previously, the neuron emits a
spike. Due to $\mu_m$, this neuron can spike even from a single input spike.}
    \label{fig:tffs_neuron}
\end{figure}

This chapter provides the basis for the hardware implementation of the SNN. The deployment of SNNs is a multi-objective hardware-software co-design effort: the SNN should provide a satisfactory classification result while allowing for fast and energy-efficient processing. An important factor determining the computational performance of SNNs is the neural code. Since recent research suggests
TTFS-coding to be the most efficient coding scheme~\cite{Guo.2021,Rueckauer.272018}, it will be used in this work. 
To implement TTFS-coding, Rueckauer et al. propose a neuron model where each neuron
can fire only a single spike~\cite{Rueckauer.272018}. This is implemented such that a neuron that has already fired cannot fire again. Note that inference on a single
sample requires multiple forward passes of the SNN. After inference on a sample is is done, the
entire SNN is reset so that all neurons are able to fire again. This only-spike-once scheme
requires a new neuron model since a single spike has to be enough to push a neurons membrane potential above firing threshold. The main modification is the introduction of the membrane potential slope $\mu_m$. This results in a the following system, illustrated in Fig. \ref{fig:tffs_neuron}:
\begin{equation}\label{ttfs_memslope}
\mu_{m_{j}}^{l}(t)= \mu_{m_{j}}^{l}(t-1)+\sum_{i} w_{i, j} \cdot x_{i}^{l-1}(t-1)
\end{equation} 
\begin{equation}\label{ttfs_mempot}
V_{m_{j}}^{l}(t)= \mu_{m_{j}}^{l}(t-1)+V_{m_{j}}^{l}(t-1)
\end{equation} 
A neuron output activation $x$ evaluates to 1 if the threshold $V_t$ is crossed and the last time step that a spike as been fired $t_{\text{spike}}$ is at 0 indicating that no spike has been emitted so far. 
\begin{equation}\label{ttfs_thresh}
x_{j}^{l}(t)=\left\{\begin{array}{c}
1 \text { if } V_{m_{j}}^{l}(t)>V_{t} \text{ and } t_{\text{spike}} = 0\\
0 \text { otherwise }
\end{array}\right.
\end{equation}

While this only-spike-once methodology results in a very high spike sparsity, it has two inherent
downsides:
\begin{itemize}
    \item The real-valued membrane potential slope has to be stored for each neuron,
effectively doubling the memory requirements of the SNN.
\item While $\mu_m$ must only be updated when spikes occur, $V_m$ of every neuron that has
received a spike must be updated every algorithmic time step as a function of $\mu_m$.
\end{itemize}

These downsides eliminate a lot of advantages of TTFS encoding. To mitigate this, Han and Roy~\cite{Han.2020} propose a modified TTFS scheme (further referred to as m-TTFS) that gets rid of $\mu_m$ at the cost of a few more spikes. The idea of \mbox{m-TTFS} is that once a neuron has exceeded $V_t$ , it emits a spike \emph{every} algorithmic time step. After a sample has been processed for $T$ time steps, the entire SNN is reset and all neurons can fire again. The m-TTFS code leads to the same number of addition operation as the standard TTFS code, but has the advantage that $\mu_m$ is not required anymore~\cite{Han.2020}.
While this m-TTFS encoding reduces spike sparsity compared to standard TTFS-encoding, it is still much more efficient than rate coding
because it reduces the number of forward passes that are required for a single inference by
orders of magnitude~\cite{Han.2020}.

Spikes are inherently discontinuous and non-differentiable. This renders the well established gradient-based backpropagation learning approach used for standard NNs impossible. The training of SNNs is still a very active research are with no established ``best practice''. Three major approaches can be identified:
\begin{itemize}
	\item \textbf{Spike Timing Dependent Plasticity}: STDP describes a large set of bio-inspired learning rules   that share one property: the synaptic weights are adapted depending on the firing time difference between neurons in layer $l$ and its preceding layer $l-1$. The basic principle is that the connection strength (=weight) of neurons is increased when they fire together, and that the connection strength is decreased when the time difference between spikes is large~\cite{Tavanaei.2019}. The standard STDP algorithm is unsupervised, i.e. no labeled training data is required.
	
	\item \textbf{Backpropagation}: this class of approaches try to overcome the non-differentiable nature of SNNs in order to enable spike-based backpropagation. Most approaches (a) try to approximate the derivatives to make gradient learning possible~\cite{Lee.2020}, (b) employ differentiable surrogate activations to replace the non-differentiable threshold operation~\cite{Neftci.2019} or (c) use a tandem approach where a SNN is performing the prediction and an equivalent standard NN is adapting the weights using backpropagation~\cite{Wu.2019}.
	
	\item \textbf{Conversion}: here the idea is  to circumvent the backpropagation problem by training a standard ANN and reusing the trained weights for the SNN. This has the major advantage that the already very mature techniques and toolchains established for standard NNs can be used for training. Deep SNNs trained with conversion methods show the best classification accuracy of all methods discussed here~\cite{Sengupta.2019, Rueckauer.2017}.
\end{itemize}
Here, we use a conversion approach to train the CSNN (see Section \ref{sec:results} for details).

\section{Hardware Design}\label{sec:Hardware_design}
The main observation is that existing implementations employ large arrays of PEs that are then highly underutilized due to spike sparsity. Another common problem is determining the number of PEs for a CNN accelerator (be it spiking or non-spiking). Most implementations like SIES~\cite{Wang.2020} or ASIE~\cite{Kang.2020} deploy a PE array where its ideal size is the size of the 2D fmap that has to be processed. However, the fmap sizes in CNNs change from model to model and from layer to layer. This fact makes maximizing the efficiency of such PE arrays very difficult. This problem of diminishing dimensions is discussed in more detail by Chen et al.~\cite{Chen.EyerissV2}. We argue that one dimension is fixed for the most parts of established CNN architectures: the kernel size. Szegedy et al. argue that convolutions with filters sizes larger than $3 \times 3$ ``might not be generally useful as they can always be reduced into a sequence of $3 \times 3$ filters''~\cite{Szegedy.2015}. Simonyan and Zisserman~\cite{DBLP:journals/corr/SimonyanZ14a} argue that, for example, two $3 \times 3$ convolutional layer have more discriminative power than a single $5 \times 5$ layer as the former incorporates two non-linearities while the latter only includes one. 
Thus, it comes with no surprise that established, well performing architectures like ResNet~\cite{He.2015}, \mbox{Inception-V3}~\cite{Szegedy.2015} or MobileNet~\cite{Howard.2017} all deploy $3 \times 3$ kernels for the vast majority of their architecture. For this reason, the proposed architecture is optimized for $3 \times 3$ kernels while $1 \times 1$ kernels for pointwise layers are also possible. Nevertheless, the techniques discussed here can also be generalized to other kernel sizes.
The core idea of this architecture is to employ less PEs but to constantly keep them busy. To ensure that the PEs run at maximum capacity, spikes are represented as address events that are compressed into queues. As soon as all spikes of a queue are processed, the next queue is selected. This allows the processing time to scale with the number of spike events and results in self-timed execution of the SNN. We start by providing a top-level overview of the hardware architecture and then proceed to show how it can be implemented efficiently on either FPGAs or ASICs.

\subsection{Top-level overview}
\begin{figure}[t!]
  \begin{center}
    \includegraphics[width=0.7\linewidth]{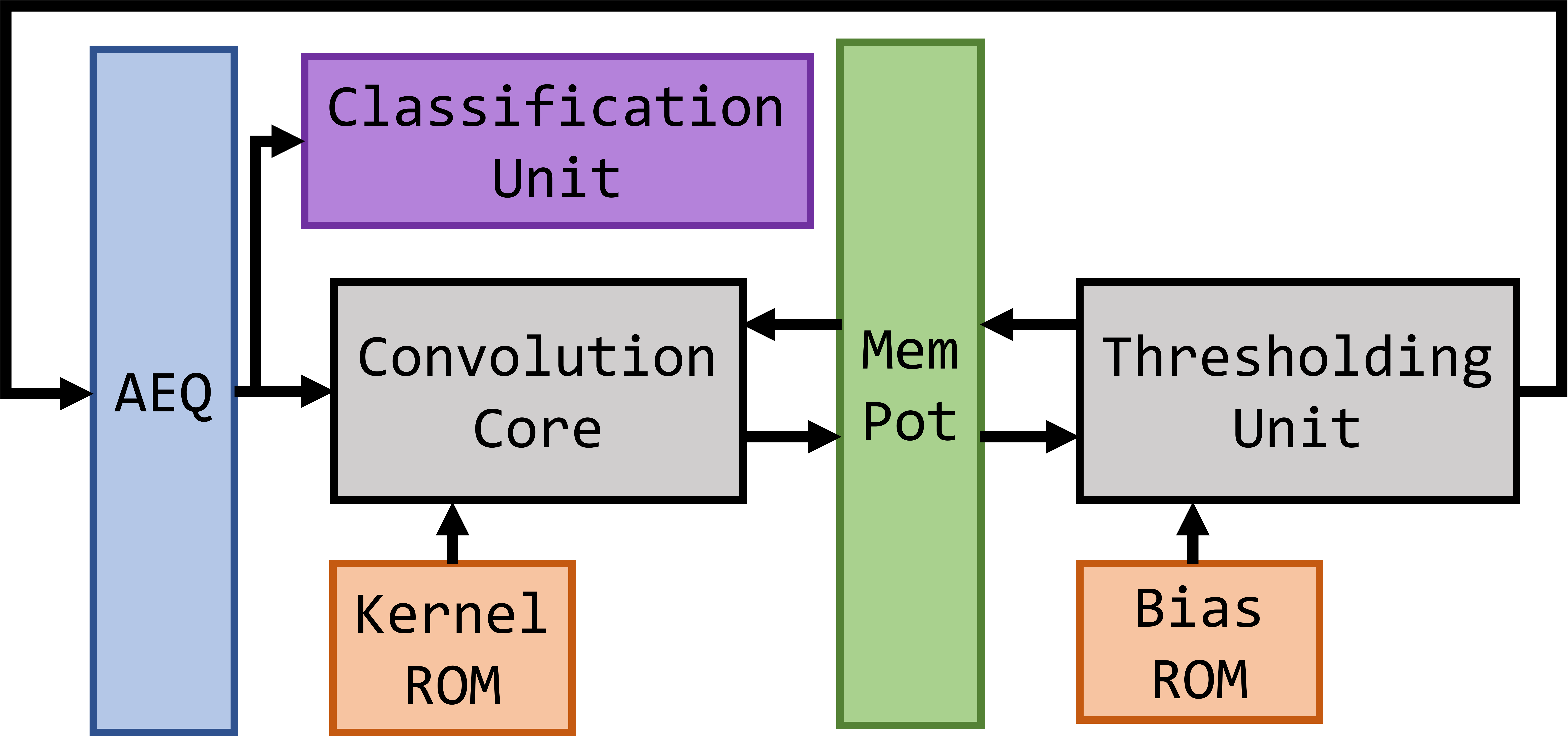}
  \end{center}
  \caption[Top-level architecture of the proposed hardware implementation.]%
  {Top-level architecture that shows the data flow between the different units. The AEQ stores the address events that are read by the convolution core. The convolution core updates the MemPot memory depending on the address events. The thresholding unit adds the bias to the neurons stored in MemPot. Also, the thresholding unit thresholds the neurons to generate the address event that are stored in the AEQ. The classification unit performs implements an Fully Connected layer and performs the final classification.}
  \label{fig:Main_architecture}
\end{figure}

The goal is to maximize the utilization of PEs while minimizing the number of PEs. To achieve this, the individual fmaps are stored in a compressed format. This compressed format is a queue of all spikes in a fmap whereby the spikes are not represented by a logical 1 but by their Address Event Representation (AER). The AER of a spike is simply the spike’s $(i, j)$ coordinates in the 2D fmap (see Fig~\ref{fig:AER_conv} for a visual example). To process an fmap, the Address Event Queue (AEQ) must be processed, which has the advantage that the number of processing steps scales directly with the number of spikes. The architecture proposed here consists of six distinct units:
\begin{itemize}
    \item The AEQ that stores the address events.
    \item The Membrane Potential memory (MemPot) for storing the $V_m$ of the neurons.
    \item The convolution unit that receives the address events from the AEQ and updates MemPot.
    \item The thresholding unit that perform multiple tasks: 
    \begin{enumerate}
        \item Threshold the MemPot and write the resulting address events to the AEQ.
        \item Perform max-pooling if required.
        \item Apply the bias to neurons in MemPot and set them back to 0 if required.
    \end{enumerate}
    \item The Read Only Memory (ROM) for storing the kernel weights $K$ and biases $b$.
    \item The classification unit performs the final classification using a small fully connected layer. Its functionality will be omitted here as the focus of this work is on accelerating the convolutional layers.
\end{itemize}
Fig. \ref{fig:Main_architecture} provides an overview over the dataflow of the architecture.

\subsection{Performing convolution with address events}
\begin{figure}[t!]
  \centering
  \includegraphics[width=0.99\linewidth]{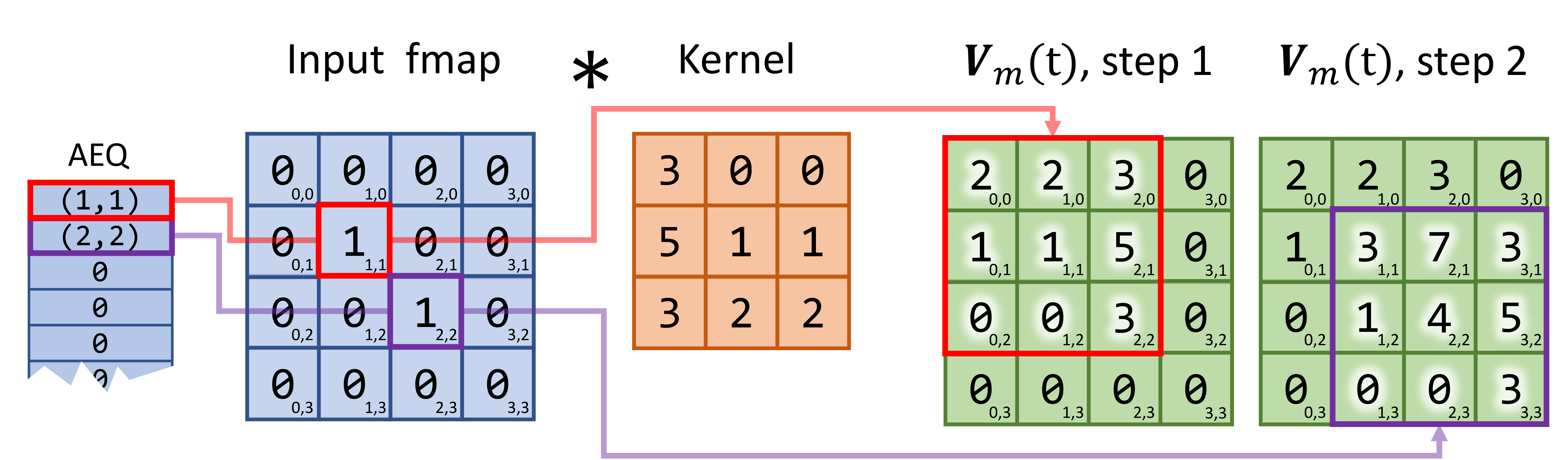}
  \caption{Convolution with address events. A 2D input fmap is stored as a queue of address events (the AEQ). To perform convolution, the AEQ is processed sequentially. Two address events allow convolution to be performed in two successive steps, whereas sliding window based convolution in this example would require $4 \times 4 = 16$ steps, one step for each pixel position in the input fmap. For each address event, 9 neurons (highlighted in white) can be updated in parallel, due to the $3 \times 3$ neighbourhood of the kernel. In $\mathbf{V}_m$, step 1 it is easy to see how the membrane potentials are updated with the kernel contents rotated by updated by $180^{\circ}$. This rotation is necessary to achieve the same result as with sliding-window based convolution (see~\cite{Morales.2019}). Note that $\mathbf{V}_m$ is initialized with zeroes. Also, note that only the AEQ is used to store spikes, the 2D Input fmap is only shown to allow for an easy interpretation of the AEQ's content.}
    \label{fig:AER_conv}
\end{figure}

Performing convolution with address events requires rethinking the well-known sliding window frame-based convolution. For this, Morales et al.~\cite{Morales.2019} propose an algorithm for event-based convolution. To process an address event at position $(i,j)$, all neurons affected by this address event have to be updated with the respective kernel weights.
The affected neurons are determined by the neighbourhood of the convolution kernel. A $3 \times 3$ kernel requires updating the neuron potentials at position $(i,j)$ and all 8 neighboring membrane potentials. To get the same result as with standard sliding-window based convolution, the respective weights of the kernel can be added to the neuron potentials by rotating the kernel by $180^{\circ}$. This is further explained in Fig. \ref{fig:AER_conv} and in~\cite{Morales.2019}. Note that this leads to the same result as performing standard sliding window based convolution. However, with this principle, only additions are needed and the number of operations scales with the number of address events in the AEQ.

Because all spike events are located inside the AEQ, the clock cycles required to perform the convolution scale directly with the number of spikes, i.e., one clock cycle per event. Due to a high degree of spike sparsity, this leads to a significant speed-up. All 9 neuron membrane potentials can be updated in parallel because there is no data dependency. Thus, a total of 9 PEs are required in the convolution unit. Each PE implements an adder that receives a membrane potential and a kernel $K[\cdot]$ and returns an updated membrane potential. Note that $K$ refers to \emph{all} kernels of the SNN, thus for each convolution the correct kernel must be selected depending on the current layer $l$, the current input channel $c_{\text{in}}$ and output channel $c_{\text{out}}$. We use the following notation to indicate the selection of the correct kernel for the current convolution: $K[c_{\text{out}}, c_{\text{in}}, l]$. Additional adders are required to calculate the addresses of the affected neurons. 

\subsection{Thresholding Unit}
\begin{figure}[!t]
  \centering
  \includegraphics[width=0.99\linewidth]{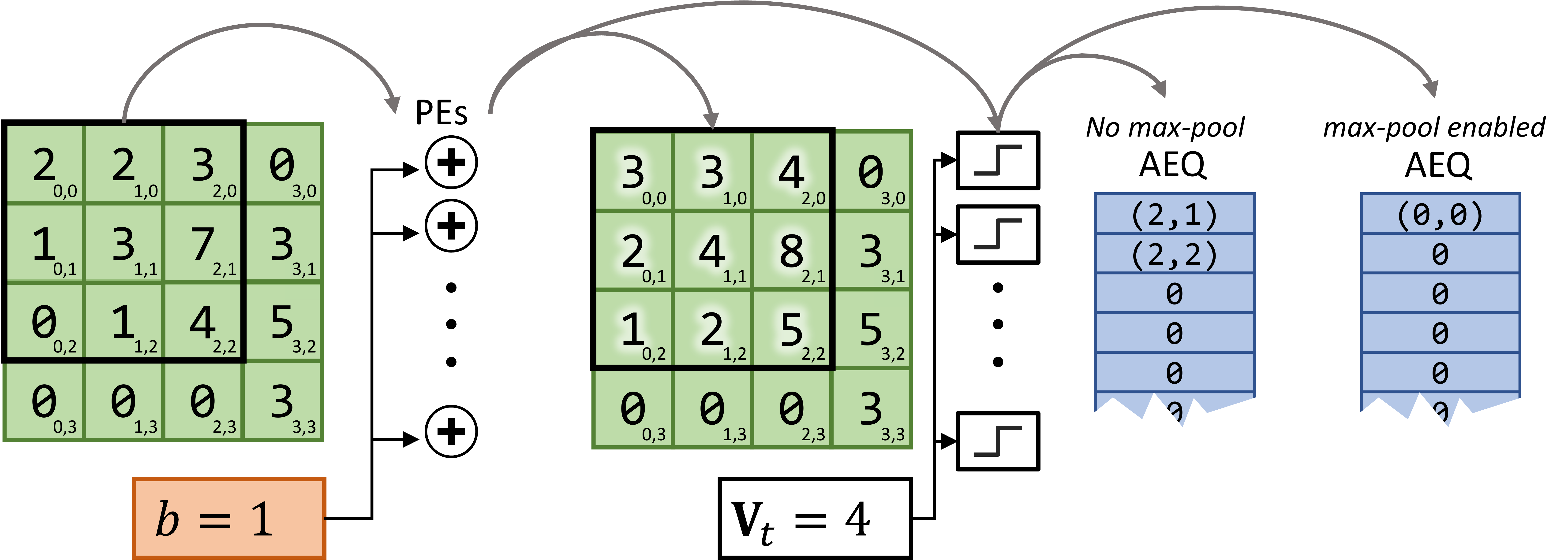}
  \caption{Functionality of the thresholding unit at the example of a single channel. Nine parallel adders add the bias and 9 parallel comparators  do the thresholding in the $3 \times 3$ window. The window is moved with a step size of three over all neurons. How max-pooling with address events works will be discussed in more detail later.}
    \label{fig:ThreshUnit}
\end{figure}

The thresholding unit performs three distinct tasks: max-pooling, thresholding and the
addition of the bias.
Like the convolution unit, the thresholding unit operates in a $3 \times 3$ neighborhood and thus processes 9 inputs in parallel. The thresholding unit does not operate in an event-based manner because \emph{all} neurons need to be visited in order to threshold them and to add the bias to them. 
The processing sequence of the thresholding unit is as follows (also illustrated in Fig.~\ref{fig:ThreshUnit}):
\begin{enumerate}
	\item Add the scalar bias $b$ to all neuron potentials in the current $3 \times 3$ window.
	\item Threshold all neuron potentials in the $3 \times 3$ window with~$V_t$:
     \begin{LaTeXdescription}
     \item[No Max-pool:] Write the addresses of the neurons in the current window to the AEQ if they exceed the threshold value $V_t$.
     \item[Max-pool enabled:] If \emph{any} neuron potential in the window crossed $V_t$, write the max-pooled address to the AEQ. How the max-pooled address can be calculated will be described later.
     \end{LaTeXdescription}
     \item The $3 \times 3$ window is moved with a stride of 3, i.e. it moves 3 pixels ahead.
     \item Repeat until all neurons of the current channel have been thresholded.
\end{enumerate}

\subsection{Scheduling strategy}
The convolution unit can only perform convolution on a single channel. To process a multichannel convolutional NN with mutiple layers, the convolution unit must be applied to all channels and
all layers in the correct order. This is a complex scheduling problem: which kernels and
biases need to be applied to which address events in which order. The naive solution to
this scheduling problem would be to implement as many convolution units as required to
perform all operations in parallel. This is not a feasible solution because hardware and
memory resources are limited.
The goal of the processing strategies proposed here is to keep the required memory resources as low as possible.
There are three types of data structures in SNNs:
\begin{itemize}
	\item The membrane potentials $\mathbf{V}_m$ stored in MemPot are not sparse and require significant memory resources.
	\item The sparse output activations are stored in the AEQ as address events.
	\item The weights and biases are stored in uncompressed form.
\end{itemize}

The key to reducing memory requirements is to reuse, i.e. multiplex, the MemPot memory for each channel.
Consider the following example: a layer has 32 channels. Each channel requires 10 kb to store the membrane potentials. Thus, 320 kb of memory are required. However, if each channel is processed one after the other, only 10 kb are required in total.
Here, an SNN is processed layer by layer. Each layer of an SNN needs to be simulated for multiple time steps $T$. To maximize the reuse of MemPot, processing is done in a channel-wise fashion. Each output channel of a layer is simulated for all time steps $t$, one channel after the other.

The output fmap of each channel is represented by its own AEQ. These AEQs can be implemented in a single dual-port RAM since each individual AEQ is processed sequentially. 
Algorithm \ref{alg:dataflow} shows the dataflow\footnote{The dataflow for standard convolution is shown. The scheme can easily be adapted to support depthwise convolution as well. For didactic reasons and lack of space, we refrain from showing this.} of the processing scheme where the output address events of a layer $l-1$ are processed to get the output address events of $l$. We use $C_{l-1}$ and $C_{l}$ to denote the number of input and output channels of the current layer $l$. The memory MemPot for storing $\mathbf{V}_m$ is only large enough for a single channel. MemPot is reused for every output channel in order to keep the memory requirements low. 

\begin{algorithm}[!t]
\small
\caption{Schematic dataflow for processing a layer $l$}
\label{alg:dataflow}
    \begin{algorithmic}[1]
    
        \For{$c_{\text{out}} \gets 0$ to $C_l$}
            \State $\mathbf{V}_m \gets 0$ \Comment{reuse for each output channel $c_{\text{out}}$}
            \For{$t \gets 0$ to $T$} \Comment{Simulate layer for all time steps $T$}
                \For{$c_{\text{in}} \gets 0$ to $C_{l-1}$}
                    \par  \Comment{Update $\mathbf{V}_m$ using the address events from the AEQ}
                    \State $\mathbf{V}_m \gets$ ConvolutionUnit(AEQ[$c_{\text{in}}$, $l-1$, $t$], 
                    \par \phantom{for fo$\mathbf{V}_m \gets$ ConvolutionUnit(} $K$[$c_{\text{out}}$, $c_{\text{in}}$, $l$], $\mathbf{V}_m$)
                  
                \EndFor
                \par  \Comment{Save the address events from the thresholding unit}    
                \State AEQ[$c_{\text{out}}$, $l$, $t$] $\gets$ ThreshUnit($b$[$c_{\text{out}}$], $V_t$, $\mathbf{V}_m$)
            \EndFor  
        \EndFor   
    \end{algorithmic}
\end{algorithm}

\section{Implementation}\label{sec:Implementation}
\begin{figure}[!t]
  \centering
  \includegraphics[width=0.99\linewidth]{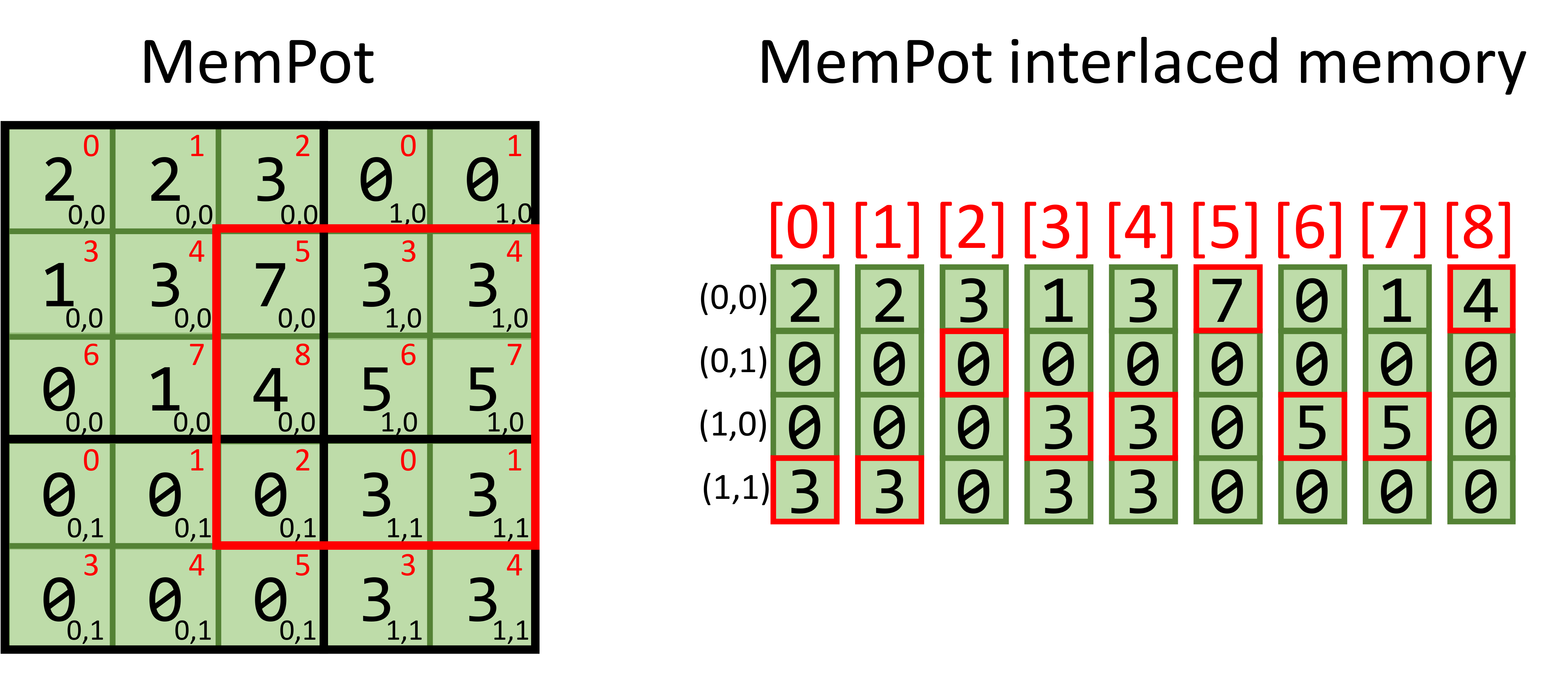}
  \caption{Memory interlacing scheme (right) and its visual representation (left). All membrane potentials are stored in the 9 memory columns 0 to 8. The elements are distributed in such a way that a $3 \times 3$ window (red) always accesses all memory columns in parallel, no matter where it is placed. Each element is uniquely addressed by its address $(i,j)$ and its $s \in (0,...,8)$ displayed in red. For example, the top left memory potential with the value 2 and the address (0,0)[0] is stored in column 0 at address 0,0. Each column can be implemented with a single dual-port RAM.}
    \label{fig:mempot_interlaced}
\end{figure}

This section shows how the top-level architecture described in the previous chapter can be
mapped to actual hardware. The top-level architecture poses one essential challenge that needs to be addressed: the membrane potential $\mathbf{V}_m$ is stored in a single memory (MemPot). However, to perform convolution or thresholding, $3 \times 3$ membrane potentials need to be accessed in parallel. Dual-port RAMs (such as BRAMs present on most FPGA chips) typically only supports one write
and one read access per clock cycle, rendering 9 parallel read/write operations impossible. To solve this problem, a novel memory distribution strategy called memory interlacing is
presented here. The idea of memory interlacing is to distribute all elements of MemPot
over 9 different RAMs called columns, such that 9 concurrent read/write operations are possible. The elements of the membrane potential need to be placed into the 9 memory columns in a 
certain fashion: regardless on which position $(i,j)$ the $3 \times 3$ window is placed, all 9 elements must
come from one memory column each. This memory interlacing scheme is further explained in Fig \ref{fig:mempot_interlaced}. Each element is addressed uniquely by its address $(i, j)$ and its column $s \in (0, ...,8)$. For a more compact notation, $(i,j)[s]$ is used to define the unique address of a neuron.
The memory interlacing scheme has multiple positive effects:
\begin{itemize}
    \item Instead of of one large monolithic memory, $\mathbf{V}_m$ is distributed to 9 smaller RAMs.
Smaller RAMs tend to be faster and more energy efficient.
\item The multiple small RAMs can be distributed closer to the PEs, which further
increases speed and energy efficiency.
\item The interlaced processing effectively prevents data-hazards.
\end{itemize}
In the following, we will discuss the hardware implementation of the different modules in detail.

\subsection{Address Event Queue (AEQ)}
\begin{figure}[!t]
  \centering
  \includegraphics[width=0.99\linewidth]{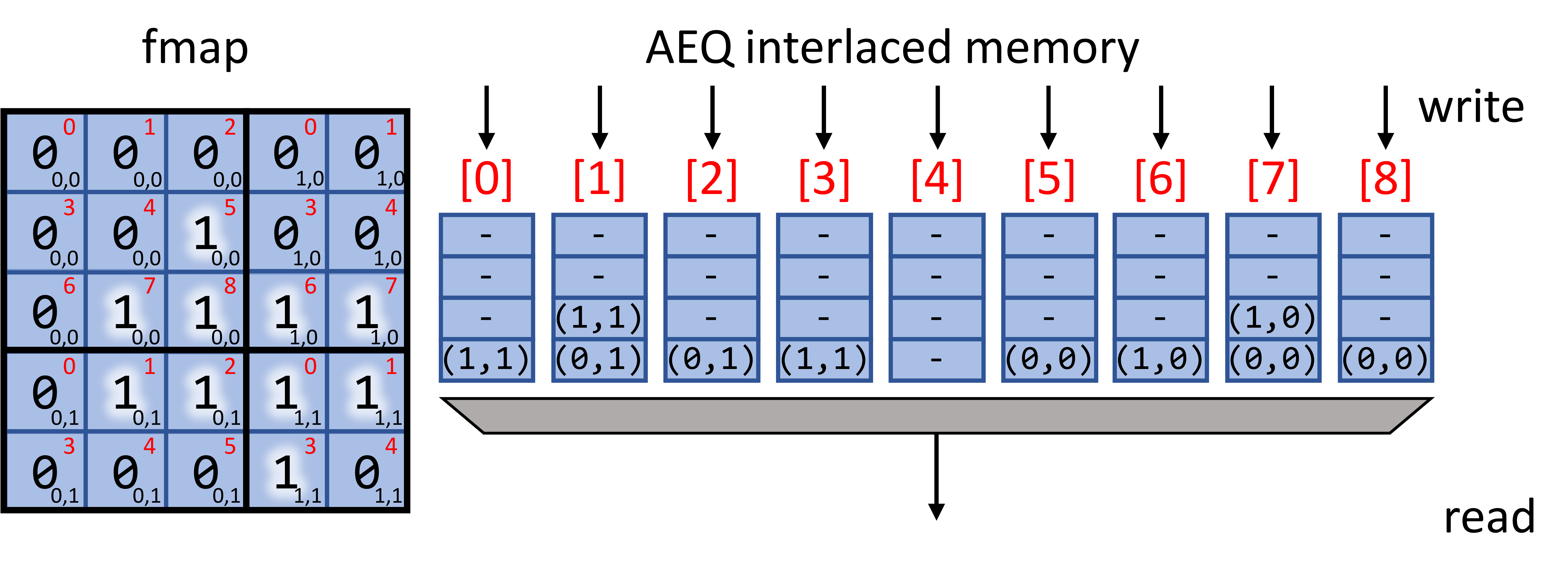}
  \caption{Interlacing scheme applied to the AEQ. The spikes represented as address events $(i,j)$ and are sorted into the respective column $s \in (0,...,8)$, displayed in red. Note that the binary fmap containing the spikes is only shown as a easy to understand visualization of the AEQ's contents.}
    \label{fig:AEQ_interlaced}
\end{figure}

The AEQ has to store the address events in queues. This is done in an interlacing fashion  (see Fig. \ref{fig:AEQ_interlaced}). The AEQ is not only a data structure but also features two independent circuits: one for writing and one for reading the queue columns. The thresholding unit writes the address events and the convolution unit reads them.

\subsubsection*{Write Logic}
The queues in the 9 columns can be filled in parallel. The 9 parallel write accesses are necessary since thresholding is performed in a $3 \times 3$ window. The write logic features 9 write counters, one for each column. An address event is only written to a queue if the respective write enable is set.

\subsubsection*{Read Logic}
Since the address events are processed one after the other, the queues are read sequentially, from queue 0 to 8.  Therefore, one read counter and the column-select counter $\in (0,..,8)$ to select the correct queue is required. Each entry in the queue not only contains the address event but also two extra bits: the valid bit and the end-of-queue bit. This valid bit indicates if an address event is valid. The valid bit is used to indicate empty queues. The end-of-queue bit indicates the last element of the queue and leads to an increment of the column-select counter. If one queue columns is completely empty, then one clock cycle is wasted by reading an invalid address event and incrementing the column-select counter. 

\subsection{Convolution Unit}
\begin{figure}[!t]
  \centering
  \includegraphics[width=0.99\linewidth]{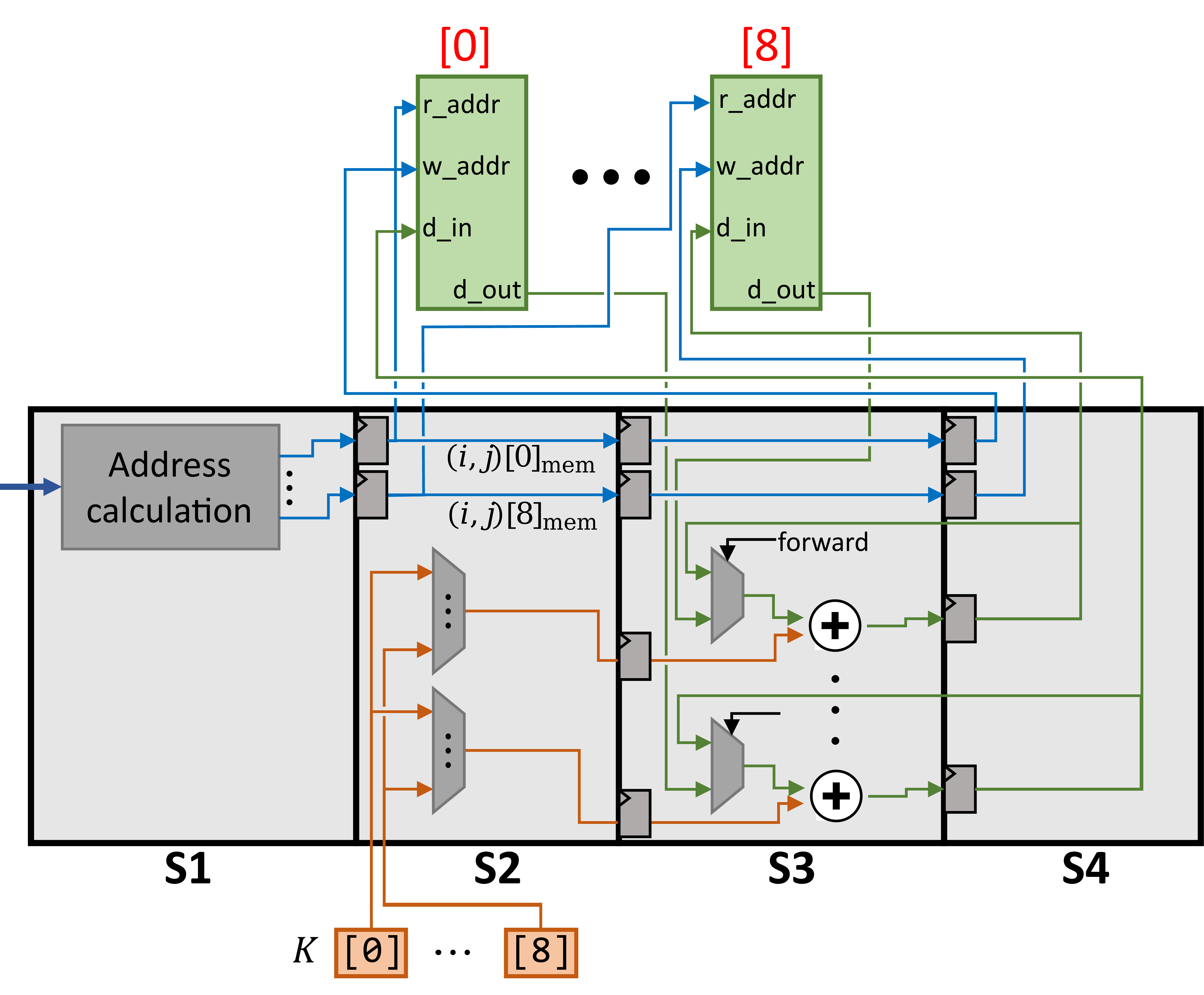}
  \caption{Pipelined architecture of the convolution unit. The convolution unit is setup in four stages. The core is operating on 9 membrane potentials (green) and 9 kernel weights (orange) in parallel. In S1, the read addresses of all 9 membrane potentials is calculated. These read addresses get passed along the pipeline and become the write addresses in S4. In S2, the read addresses are delivered to the memory columns. It takes one clock cycle for the MemPot memory to deliver the output data. Also, the 9 kernel elements (orange) are sorted in the right order and transported to the PEs in S3. The adders in S3 then update the membrane potentials. In S4, the updated membrane potentials are written back. Note how each PE is permanently connected to the respective memory column of MemPot. For better visibility, only the logic and data paths of two PEs are displayed, the dots $\cdot \cdot \cdot$ indicate the existence of the missing 7 data paths.}
    \label{fig:Pipelined_ConvCore}
\end{figure}

The functionality of the convolution unit can be split into multiple sub-units: address
calculation, kernel permutation, MemPot update calculation and data hazard detection.
The circuit of the convolution unit is pipelined into four stages S1 to S4. The general data
flow in the convolution unit is as follows:
\begin{LaTeXdescription}
	\item[Start] Receive address event: When the AEQ receives a read-enable signal, it starts to read out the 9 queues sequentially from a given offset. The convolution unit receives the input address event $(i,j)[s]_{\text{in}}$ to uniquely identify the location of an incoming spike.
	\item[S1] Calculate addresses: The convolution unit calculates the addresses of all affected neurons in the $3 \times 3$ neighborhood. 
	\item[S2] Read MemPot: The convolution unit reads the 9 membrane potentials from the calculated addresses. Also, the permutated kernel will be selected here.
	\item[S3] Calculate update: The convolution unit adds the respective kernel weights to the 9 membrane potentials.
	\item[S4] Write back MemPot: The 9 updated membrane potentials are written back to MemPot to the same addresses calculated in S1. 
\end{LaTeXdescription}
Pipelining increases the parallelism inside the convolution unit because all four stages
are executing their respective task in parallel (see Fig. \ref{fig:Pipelined_ConvCore}). This ensures that all parts
of the convolution unit are busy. For example, S3 is updating the membrane potentials
of an address event while S2 is already fetching all membrane potentials for the next
address event. Furthermore, since the combinational parts of the circuit are divided into
four stages, a much higher clock frequency can be achieved. However, pipelining also has
some downsides. At the start, it takes four address events until the pipeline is completely
filled (wind-up). Only a full pipeline can deliver maximum efficiency. It is therefore
important that a constant flow of address events is provided so that the pipeline stages are always saturated. This constant supply can be provided since the address events of a channel
are compressed into a queue that is read event by event. A downside of pipelining is the occurrence of \emph{data hazards} caused by data dependencies.

\subsubsection*{Address calculation}
The convolution unit receives an input address event $(i,j)[s]_{\text{in}}$ from the AEQ. The address calculation logic must then compute what the addresses of the nine affected membrane potentials in the kernel neighbourhood are so that they can be fetched from MemPot. Here, we use the subscript ``in'' to refer to the components of the input address events and ``mem'' to refer to the components of a neuron's address in MemPot. 
Nine different addresses $(i,j)_{\text{mem}}$ need to be calculated for each of the 9 MemPot memory columns: $(i,j)[0]_{\text{mem}},...,(i,j)[8]_{\text{mem}}$. 
Consider the following example: an input address event has the address $(i,j)[s]_{\text{in}}$. 
For simplicity, consider only the $(i,j)_{\text{mem}}$ addresses for column 0 ($(i,j)[0]_{\text{mem}}$). If the input address event comes from column $s_{\text{in}} \in \{2,5,8\}$ of the AEQ, then $i_{\text{mem}} = i_{\text{in}}+1$ otherwise $i_{\text{mem}} = i_{\text{in}}$ as described in Eqn.~\eqref{eq:column_0i}. Similarly, Eqn.~\eqref{eq:column_0j} describes how $j_{\text{mem}}$ can be calculated. For the remaining columns $s_{\text{mem}}$, the address calculation logic can be constructed in a similar fashion, which we refrain from showing here due to spacial limitations.

  \begin{equation}\label{eq:column_0i}
    i_{\text{mem}}=
    \begin{cases}
      i_{\text{in}}+1, & \text{if} \ s_{\text{in}}\in \{2,5,8\} \\
      i_{\text{in}},   & \text{otherwise}
    \end{cases}, \text{for} \ s_{\text{mem}}=0
  \end{equation}
  
    \begin{equation}\label{eq:column_0j}
    j_{\text{mem}}=
    \begin{cases}
      j_{\text{in}}+1, & \text{if} \ s_{\text{in}}\in \{6,7,8\} \\
      j_{\text{in}},   & \text{otherwise}
    \end{cases}, \text{for} \ s_{\text{mem}}=0
  \end{equation}

\begin{figure}[!t]
  \begin{center}
    \includegraphics[width=0.5\linewidth]{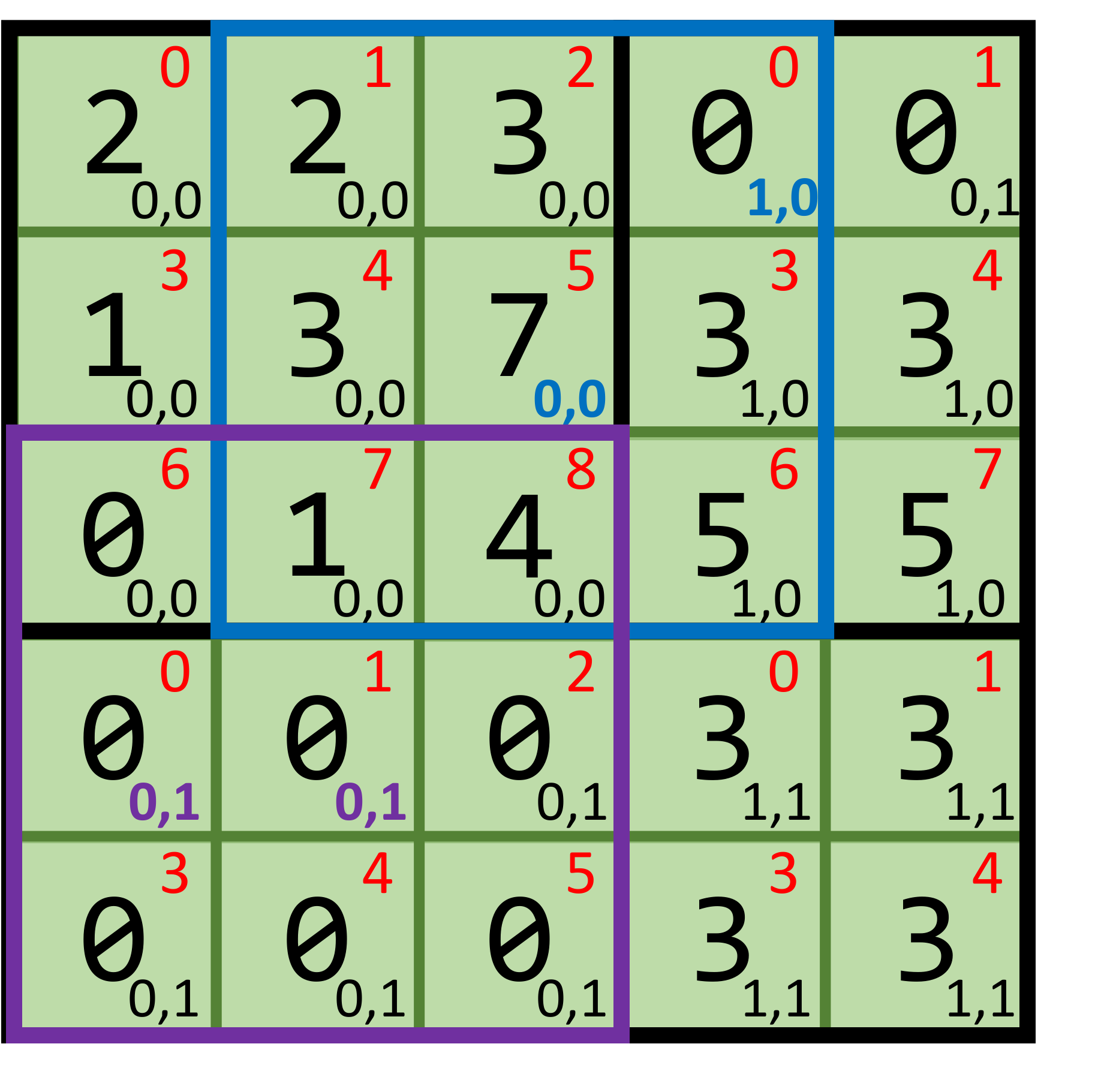}
  \end{center}
  \caption[Example for the address calculation scheme with interlaced memory.]%
  {Example for the address calculation scheme with interlaced memory. Blue kernel: the input address event $(0,0)[5]_{\text{in}}$ comes from $s_{\text{in}} = 5$, which defines the center of the kernel. To calculate the address $i_{\text{mem}}$ for $s_{\text{mem}}=0$, Eqn.~\eqref{eq:column_0i} is used to get $i_{\text{mem}}=i_{\text{in}}+1=1$. Purple kernel: the input address event is $(0,1)[1]_{\text{in}}$, defining the center of the kernel. To calculate the $i_{\text{mem}}$ for $s_{\text{mem}}=0$, Eqn.~\eqref{eq:column_0i} is used to get $i_{\text{mem}}=i_{\text{in}}=0$ since $s_{\text{in}}=1$ is not 2 or 5 or 8.}
  \label{fig:address_calc_example}
\end{figure}

This example for $i_{\text{mem}}$ is described visually in Fig.~\ref{fig:address_calc_example}. Four adders are needed to calculate $i+1, i-1, j+1, j-1$ and 9 comparators are required to check from which of the 9 columns in the AEQ the address event came from. Parts of the kernel can be out of bounds of the fmap when the kernel center is directly at the fmap's edge. To avoid errors, this out-of-bound condition must be detected so that no membrane potential updates occur for the parts in question. Out-of-bounds detection is performed by detecting under/overflows in the address calculation logic. For example, an address event at $(0,0)[0]$ would cause MemPot column $s_{\text{mem}} = 8$ to be addressed with $(-1,-1)$. Since the address calculation logic only supports non-negative integers, this would cause an underflow that can be detected with very little hardware overhead.

\subsubsection*{Kernel permutation}
Each of the 9 PEs is essentially an adder, performing the updates of the membrane potentials is connected to one of the 9 memory columns of MemPot. Each PE receives on of the nine kernel elements $K[0]$ to $K[8]$ and a membrane potential as an input. While a channel is processed, the kernel itself does not change. However, the mapping of the 9 kernel elements $K[\cdot]$ to the 9 PEs changes depending on the location of the input column $s_{\text{in}} = 5$ that the address event came from. Reconsider the example in Fig. \ref{fig:address_calc_example}. For the blue kernel, the top left kernel element ($K[0]$, assuming that the kernel is already flipped by $180^{\circ}$) has to be mapped to the PE of MemPot column $s_{\text{mem}} = 1$. For the purple kernel however, $K[0]$ has to be mapped the PE of $s_{\text{mem}} = 6$. Since there are 9 memory columns there are 9 different permutations of the kernels weights. All 9 possible permutations are calculated in parallel in S2. The correct permutation is then selected with a multiplexer for each PE (see Fig. \ref{fig:Pipelined_ConvCore}).  The hardware cost is relatively low: a total of nine 9-to-1 multiplexers are needed.

\subsubsection*{Update calculation}
During the calculation of the membrane potential update, arithmetic overflows or underflows might occur. When an overflow occurs, a large membrane potential overflows and becomes a negative membrane potential. Underflows are even more critical because strongly negative membrane potentials become very large membrane potentials, generating erroneous spikes. The obvious solution to this problem would be to adapt the bit widths of all data paths so that over/underflows are impossible. However, this would significantly increase the required hardware resources. 
Instead, \emph{saturation} arithmetic is used here.   
If the result of an addition is larger than the maximum, it is clamped to the maximum representable value. In a similar fashion, too small values are clamped to the minimum representable value. Saturation works well for SNNs with m-TTFS coding: A further decrease of an already very negative membrane potential has no effect on the output of the neuron. Similarly, further increasing the membrane potential when it is well above firing threshold does not change the neurons output. To implement saturation arithmetic in hardware, over/underflow detection is required. Over/underflow detection only requires checking a single bit and thus is cheap to implement.

\subsubsection*{Data hazard mitigation}
In this architecture, Read After Write (RAW) data hazards can occur when an updated membrane potential is read from memory that has not yet been calculated or written back. When the RAW-hazard is not handled, the update of an membrane potential will be overwritten. 
There are two situations where a RAW-hazard can occur: between S2-S4 and between S2-S3.
In the case of S2-S4, S2 reads from the same address that S4 is currently writing to. Writing to the memory takes one clock cycle and thus S2 receives an outdated membrane potential. Resolving the S2-S4 hazard is cheap because the updated membrane potential has already been calculated, just not written back yet. The updated membrane potential must only be forwarded directly to S3, bypassing MemPot (see Fig. \ref{fig:Pipelined_ConvCore}).
The S2-S3 hazard occurs when S2 is reading from a memory address for which S3 is currently calculating an update. In this case, forwarding does not work because the update has not yet been calculated. The solution is to stall S2, S1 and the AEQ for one clock cycle so that S3 can finish the calculation. The S2-S3 hazard is then a S2-S4 hazard that can be resolved by forwarding. 
To implement hazard detection, 9 comparators are required for S3 and 9 comparators are required in S4. The forwarding logic is cheap to implement: only 9 2-to-1 multiplexer are required, one for each PE. 

S2-S4 hazards are not a problem because they can be resolved by forwarding. S2-S3 hazards should be avoided because they require parts of the pipeline to be stalled. This reduces throughput and leaves parts of the pipeline unoccupied. The probability of S2-S3 hazards is greatly reduced due to the design of the AEQ interlaced memory. S2-S3 hazards occur when two immediately successive addressing events access overlapping membrane potentials. See for example Fig. \ref{fig:address_calc_example}. If the address event 0,1 column 1 (purple) is  processed directly after event 0,0 column 5 (blue) then a S2-S3 hazard would occur for column 7 and 8 (where the kernels overlap). This cannot happen with the AEQ design because all address events from memory columns 2, 3 and 4 are processed first. Apart from that, the address events are read from the AEQ column-wise, i.e. first all address events from column 0 are read, then column 1 and so on. This is a major advantage because address events from the same column index will never access overlapping membrane potentials. Processing address events from the same column will never result in data hazards. Data hazards can only occur on switching from one column to another.

\subsection{Thresholding Unit}
\begin{figure*}[!t]
  \centering
  \includegraphics[width=0.7\linewidth]{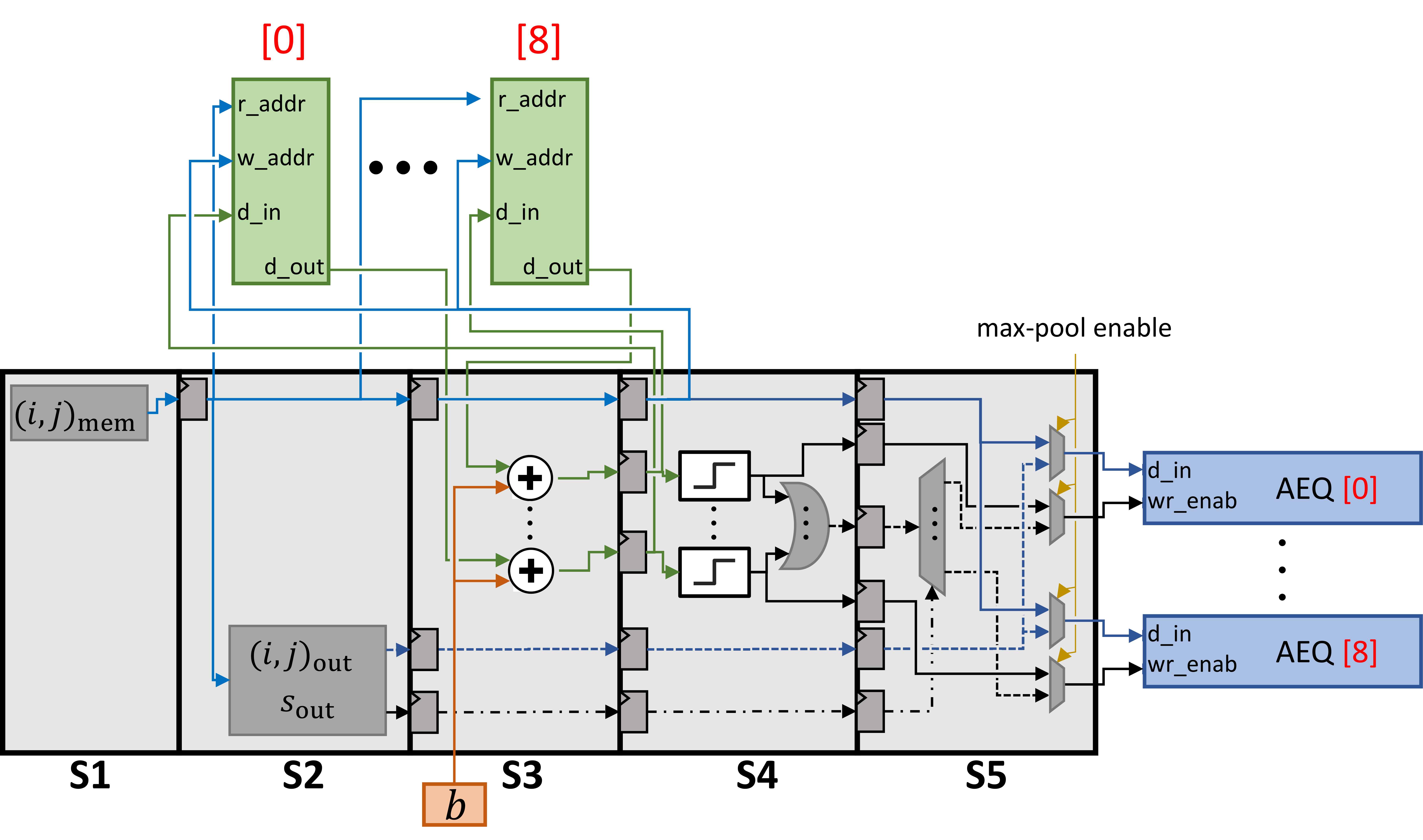}
  \caption{Pipelined thresholding unit. The MemPot addresses $(i,j)_{\text{mem}}$ are calculated in S1. In S2, the MemPot is read and the max-pooled address and memory column are calculated. In S3, the bias update is calculated. In S4, the thresholding is performed. Also, the result of the bias update is written back. For max-pooling, a 9-to-1 or-gate combines all outputs of the comparators. In S5, the results of the thresholding are written to the AEQ columns. If max-pooling is enabled, then the max-pooled address is forwarded to the data ports of the AEQ columns. The calculated max-pool column is used to select the correct AEQ column's write enable. If max-pooling is disabled, the MemPot addresses $(i,j)_{\text{mem}}$ are connected to all data ports. Also, each comparator is connected to its AEQ column via the write enable. Thus, 9 address events can be written in parallel. Signals related to max-pooling are indicated by dashed lines.}
    \label{fig:thresh_unit_pipelined}
\end{figure*}

The thresholding unit has to visit every membrane potential in the MemPot.
It does so by sliding a $3 \times 3$ window over the membrane potential with a stride of 3. 
The general architecture shows some similarities to the convolution unit. The thresholding unit starts its operation as soon as its clock enable signal is set to 1. A pipelined design with five stages S1 to S5 deployed:
\begin{LaTeXdescription}
	\item[S1] Calculate addresses: The thresholding unit calculates the addresses of all 9 membrane potentials in the $3 \times 3$ window that is currently processed.
	\item[S2] Read MemPot: The thresholding unit reads the 9 membrane potentials form MemPot.
	\item[S3] Add Bias: The scalar bias is added to all 9 membrane potentials.
	\item[S4] Threshold: The updated membrane potentials are compared to the threshold $V_t$ to determine if they fire a spike.
	\item[S5] Write MemPot and AEQ: The updated membrane potentials are written back to MemPot. If a spike is generated in S4, then the respective AEQ-column is written. Also,when enabled,  max-pooling is performed here.
\end{LaTeXdescription}
It is important to note that no data hazards can occur in the thresholding unit. This is because each membrane potential is read out only once. Fig. \ref{fig:thresh_unit_pipelined} gives a schematic overview of the thresholding unit. 

It is important to note that no data hazards can occur in the thresholding unit. This is
because each membrane potential is read out only once. 

\subsubsection*{Address Calculation}
Address calculation is very simple, thanks to the interlaced memory. Only two counters are required, one for the $i_{\text{mem}}$ coordinate and one for the $j_{\text{mem}}$ coordinate. Addressing all memory columns of MemPot with the same $(i,j)_{\text{mem}}$ address accesses a $3\times 3$ window by design (see for example Fig. \ref{fig:max_pool_interlace} where all membrane potentials with the coordinate (0,0) are located in a $3\times3$ window). 

\subsubsection*{Bias update}
The scalar bias is used as an input for all 9 PE (as illustrated in Fig. \ref{fig:ThreshUnit}). Saturation arithmetic is used to avoid underflows and overflows, akin to that of the convolution unit.

\subsubsection*{Thresholding}
Thresholding is performed by 9 parallel comparators. Due to m-TTFS encoding, each neuron that has fired already needs to fire again. This is implemented with a spike indicator bit that is stored together with the neurons membrane potential in the MemPot memory. The comparators check two conditions: if the firing threshold is crossed and if the spike indicator is set. If any of these two condition is true, then the spike indicator bit of the respective neuron is set to 1. It is only set back to 0 if a new sample has to be processed.

\subsubsection*{Writing the AEQ}
\begin{figure}[!t]
  \begin{center}
    \includegraphics[width=0.8\linewidth]{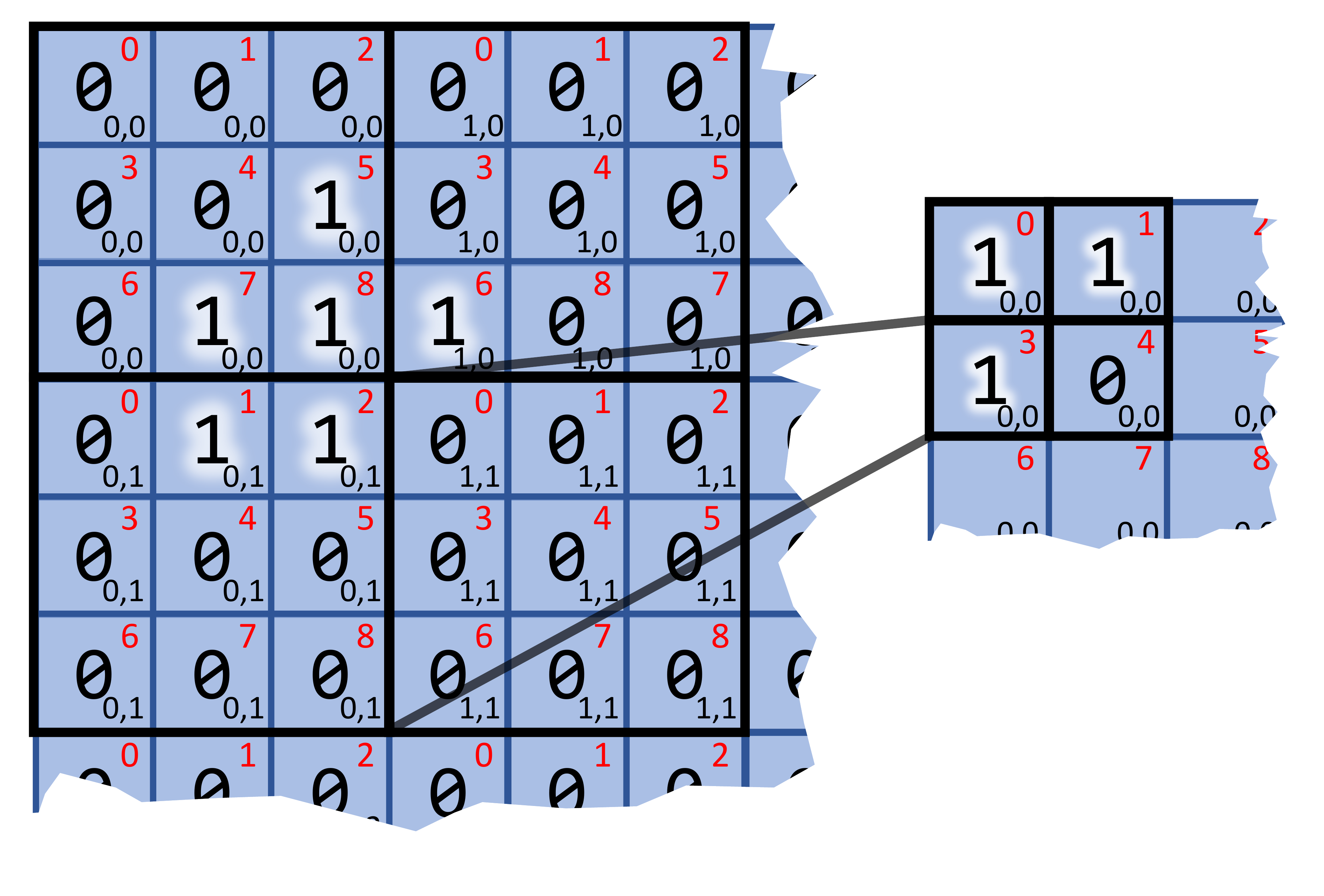}
  \end{center}
  \caption[Max-pooling with interlacing memory columns.]%
  {Max-pooling with interlacing memory columns requires the calculation of the address $(i,j)[s]_{\text{out}}$ and the respective memory column. For example, all spikes that are generated from address $(0,1)[0]_{\text{mem}}$ to $(0,1)[8]_{\text{mem}}$ are pooled to a single address event with the address $(0,0)[3]_{\text{out}}$.}
  \label{fig:max_pool_interlace}
\end{figure}

How the generated address events (indicated with a ``out'' subscript) are written to the AEQ depends on whether max-pooling is enabled or not. The simplest case is when max-pooling is disabled.
In this case, the address event $(i,j)[s]_{\text{out}}$ of a spiking neuron is simply its MemPot address $(i,j)[s]_{\text{mem}}$. With $3 \times 3$ max-pooling enabled, the resulting address event $(i,j)[s]_{\text{out}}$ has to be calculated. Consider the example in Fig. \ref{fig:max_pool_interlace}. For example, all spikes from the addresses $(0,1)[0]_{\text{mem}}$ to $(0,1)[8]_{\text{mem}}$ have to be mapped to a single address event $(0,0)[3]_{\text{out}}$. The calculation of this mapping is inexpensive to implement in hardware with four counters that run along with the address calculation logic. To avoid expensive division operations, a sequential circuit can be constructed using only adders to perform the address and column index calculation. An algorithmic representation is can be seen in Algorithm \ref{alg:max_pol_calc}.  
The counters $i_{\text{mem}},j_{\text{mem}}$ are used to calculated the addresses of the $3\times 3$ window applied to MemPot. The counters $s_{\text{out},i}$ and $s_{\text{out},j}$ are used to calculate the $s_{\text{out}}$ of the max-pooled address event. To save power, the clock enable of the max-pooling calculation logic can be turned off. 

\begin{algorithm}[!t]
  \small
  \caption{Calculate max-pooled address event}
  \label{alg:max_pol_calc}
  \begin{algorithmic}[1]

  \State $s_{\text{out},i}\gets 0$ \Comment{Counts in the sequence 0,1,2,0,1,2,...}
  \State $s_{\text{out},j}\gets 0$ \Comment{Counts in the sequence 0,3,6,0,3,6,...}
  \State $i_{\text{out}} \gets 0$ 
  \State $j_{\text{out}} \gets 0$ 
    \For{$j_{\text{mem}}\gets 0$ to $j_\text{max}$}      
        \For{$i_{\text{mem}}\gets 0$ to $i_\text{max}$}
            
            \If{$i_{\text{mem}} = i_\text{max}$} 
                \State $s_{\text{out},i} \gets 0$
                \State $i_{\text{out}} \gets 0$
                \If{$s_{\text{out},j} = 6$} 
                    \State $s_{\text{out},j} \gets 0$
                    \State $j_{\text{out}} \gets j_{\text{out}} + 1$
                \Else
                    \State $s_{\text{out},j}\gets s_{\text{out},j} + 3$
                \EndIf
            \Else
                \If{$s_{\text{out},i} = 2$} 
                    \State $s_{\text{out},i} \gets 0$
                    \State $i_{\text{out}} \gets i_{\text{out}} + 1$
                \Else
                    \State $s_{\text{out},i} \gets s_{\text{out},i} + 1$
                \EndIf 
            \EndIf
            \State  $s_{\text{out}} = s_{\text{out},i} + s_{\text{out},j}$ \Comment{Max-pooled column}
        \EndFor        	
    \EndFor
  \end{algorithmic}
\end{algorithm}

\section{Experiments and Evaluation}\label{sec:results}
To evaluate the effectiveness of the proposed architecture, we first trained a small CSNN on the MNIST\footnote{\url{http://yann.lecun.com/exdb/mnist/}} and the more difficult Fashion-MNIST\footnote{\url{https://github.com/zalandoresearch/fashion-mnist}} dataset. Training was performed with a conventional CNN using Tensorflow Keras\footnote{\url{https://keras.io/about/}} the clamped ReLU activation function (as described by Rueckauer et al.~\cite{Rueckauer.272018}). In preparation for the deployment in hardware, the CNN was then retrained using quantization-aware training~\cite{Jacob.2017}. The weights of the CNN were then converted using the SNN-Toolbox\footnote{\url{https://snntoolbox.readthedocs.io/en/latest/}} proposed by Rueckauer et al.~\cite{Rueckauer.272018, Rueckauer.2017} and quantized to 8 and 16 bit. The CSNN has a structure of \mbox{$(28 \times 28$-32C3-32C3-P3-10C3-F10)}. The notation is as follows. Convolutional layers: $<$\#channels$>$C$<$kernel~size$>$, Max-pooling layers: P$<$window~size$>$, Fully-connected layers: F$<$\#neurons$>$. Experimentally, it was found that simulating this m-TTFS encoded CSNN for $T=5$ time steps yielded the best classification accuracy.
The input frames consist of integer pixels that need to be binarized before processing in order to get input spikes that are then fed to the CSNN. The binarization can be achieved by thresholding the individual frames. For a simple dataset like MNIST where the background is clearly separated from the object, applying only a single threshold is sufficient. However, this inevitably leads to a loss of information and we thus propose to convert the input frame into binary spikes by a applying a set of thresholds $P = (p_1, p_2,...,p_{T-1})$. An important property of $P$ is that it is a strictly increasing set to mimic m-TTFS encoding. 

\begin{figure*}
     \centering
     \begin{subfigure}[b]{0.3\textwidth}
         \centering
         \includegraphics[scale=1]{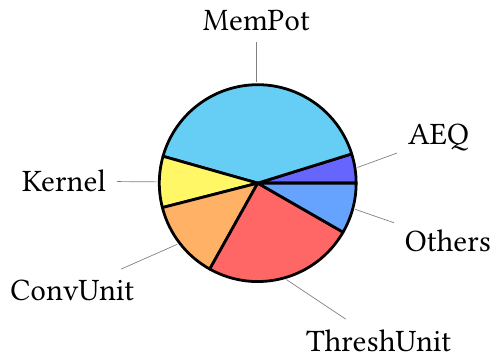}
         \caption{LUTs}
         \label{fig:LUT}
     \end{subfigure}
     \hfill
     \begin{subfigure}[b]{0.3\textwidth}
         \centering
         \includegraphics[scale=1]{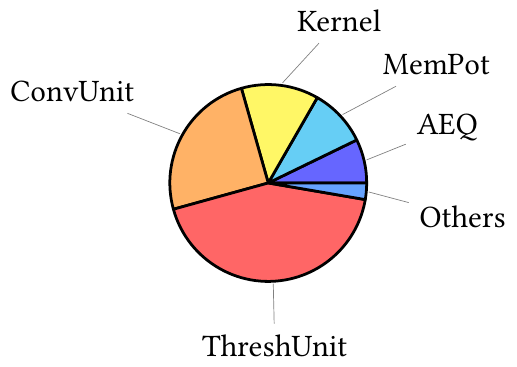}
         \caption{Flip-Flops}
         \label{fig:FF}
     \end{subfigure}
     \hfill
     \begin{subfigure}[b]{0.3\textwidth}
         \centering
         \includegraphics[scale=1]{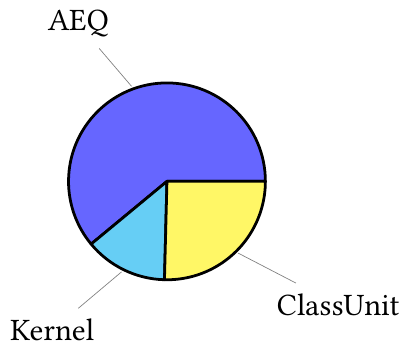}
         \caption{BRAMs}
         \label{fig:BRAM}
     \end{subfigure}
        \caption{Utilization if the different FPGA resources. ``Others'' includes the control unit, the classification unit and the bias ROM. Due to space limitations, units such as the classification unit that implements the final fully connected layer could not be described here. Note that the MemPot memory rows were too small to map efficiently to BRAM, so they were implemented as distributed LUT-RAM.}
        \label{fig:Utilization}
\end{figure*}

The proposed architecture was synthesized for the Xilinx Zynq UltraScale+ XCZU7EV FPGA. The power estimation is performed with the Vivado Power Estimator tool. The general architecture of the accelerator is very compact, therefore the parallelism and thus the latency can be improved by implementing multiple units in parallel. We tested our implementation with multiple degrees of parallelization, referring to the number of AEQs, MemPot memories, Kernel and Bias ROMs, thresholding units and convolution units. We found that for this CSNN, a parallelization of $\times 8$ yielded the best energy efficiency (see Table~\ref{tab:parallelization}). Table~\ref{tab:synthesis} provides detailed synthesis and utilization results (in terms of LUTs, Flip-Flops (FFs), Block-RAM (BRAM) and dedicated DSPs) and compares them to related work. To evaluate the  effectiveness of the proposed approach, we compared the input activation sparsity for each layer with the PE utilization in Table~\ref{tab:sparsity}. Sparsity refers to the number of non-zero activations in relation to all activations. PE utilization measures the clock cycles in which the PEs receive valid address events relative to all clock cycles required to process the CSNN. Note that the PE utilization does not take into account that there might be zero weights that do not lead to and update to MemPot. Table~\ref{tab:performance} shows the performance statistics of the architecture proposed here and compares them to other CSNN MNIST implementations. Table~\ref{tab:Fashion} compares the accuracy on the Fashion-MNIST dataset with related work.  Direct quantitative comparisons should be made with caution because the different approaches do not use identical SCNN architectures for their experiments. Nonetheless, the SCNN sizes are comparable enough (all contain three to four trainable layers) to allow for a meaningful qualitative comparison. Fig.~\ref{fig:Utilization} provides an overview over how may hardware resources are consumed by the individual units.

\begin{table}
  \caption{Performance of different degrees of parallelism, here for the 8 bit implementation. Here, parallelism refers to the number of parallel convolution cores, AEQs, thresholding units, etc.}
  \label{tab:parallelization}
  \centering
  \begin{tabular}{lccccc}
    \toprule
    Parallelization                     & $\times 1$ & $\times 2$ & $\times 4$ & $\times 8$ & $\times 16$\\
    \midrule
    Throughput [FPS]                    & 3,077       & 5,908       & 10,987     & 21,446        & 33,292 \\
    Efficiency [FPS/W]                  & 3,149       & 5,006       & 7,474      & 10,163        & 9,148   \\
  \bottomrule
\end{tabular}
\end{table}

\begin{table}
  \caption{FPGA Synthesis Results, compared to other FPGA-based SNN implementations.}
  \label{tab:synthesis}
  \centering
  \begin{tabular}{lccccc}
    \toprule
                                 & \multicolumn{1}{c}{\begin{tabular}[c]{@{}c@{}}Frequency\\ {[}MHz{]}\end{tabular}} & LUT    & FF      &   \multicolumn{1}{c}{\begin{tabular}[c]{@{}c@{}}BRAM\\ {[}Mb{]}\end{tabular}}   & DSP  \\
    \midrule 
    This work (8 bit)            & 333             & 19 k   &  12 k   & 2.1        & 32   \\
    This work (16 bit)           & 333             & 33 k   &  21 k   & 3.9        & 64   \\
    Fang et al.~\cite{Fang.2020} & 125             & 115 k  &  233 k  & 9.1        & 1.7 k   \\
    Guo et al.~\cite{Guo.2019}   & 100             & 53 k   &  100 k  & 2.3        &      \\
    SIES~\cite{Wang.2020}        & 200             & 302 k  &  421 k  & 6.9        &      \\
  \bottomrule
\end{tabular}
\end{table}

\begin{table}
  \caption{Sparsity of each layers input activations compared to the PE utilization for each convolutional layer. Here for the very first sample of the MNIST validation dataset.}
  \label{tab:sparsity}
  \centering
  \begin{tabular}{lccc}
    \toprule
    Convolutional Layer                 & Layer 1 & Layer 2 & Layer 3 \\
    \midrule
    Input activation sparsity                 & 93\%   & 98\%     & 98\%    \\
    PE utilization                      & 72\%   & 58\%     & 56\%    \\
  \bottomrule
\end{tabular}
\end{table}

\begin{table}
  \caption{Accuracy on the Fashion-MNIST dataset compared to other works.}
  \label{tab:Fashion}
  \centering
  \begin{tabular}{lccc}
    \toprule
    Work                 & This work & Guo et al.~\cite{Guo.2019} & Fang et al.~\cite{Fang.2020} \\
    \midrule
    Accuracy [\%]                 & 88.9   & 87.5     & 89.2    \\
    Quantization [bits] & 16 & 32 & 16 \\
  \bottomrule
\end{tabular}
\end{table}

\begin{table*}
\caption{Performance comparision for different computing platforms on the MNIST dataset, sorted by efficiency [FPS/W] in descending order.}
\label{tab:performance}
\centering
\begin{minipage}{\textwidth}
\begin{center}
  \begin{tabular}{lccccccc}
    \toprule
                                                                               & Type  
                                                                               &\multicolumn{1}{c}{\begin{tabular}[c]{@{}c@{}}Quantization\\ {[}bits{]}\end{tabular}}
                                                                               &\multicolumn{1}{c}{\begin{tabular}[c]{@{}c@{}}Throughput\\ {[}FPS{]}\end{tabular}}     
                                                                               &\multicolumn{1}{c}{\begin{tabular}[c]{@{}c@{}}Latency\\ {[}ms{]}\end{tabular}}
                                                                               &\multicolumn{1}{c}{\begin{tabular}[c]{@{}c@{}}Power\\ {[}W{]}\end{tabular}}  
                                                                               &\multicolumn{1}{c}{\begin{tabular}[c]{@{}c@{}}Efficiency\\ {[}FPS/W{]}\end{tabular}} 
                                                                               &\multicolumn{1}{c}{\begin{tabular}[c]{@{}c@{}}Accuracy\\ {[}\%{]}\end{tabular}}  \\
    \midrule                                                                                
    This work                                                                                                 & FPGA  & 8  & 21 k   &  0.04        & 2.1        & 10163              & 98.3           \\
    
    This work                                                                                                 & FPGA  & 16  & 21 k   &  0.04        & 2.9        & 7208               & 98.2           \\
    Fang et al.\textsuperscript{\ref{note1}}~\cite{Fang.2020}                                                  & FPGA  & 16 & 2124   &  0.52        & 4.5        & 471                & 99.2       \\
    Loihi\textsuperscript{\ref{note1}}~\cite{Davies.2018}                                                      & ASIC  &    & 671    &  1.5         & 3.8        & 178                & 98.0       \\  
    Jetson\textsuperscript{\ref{note1}}                                                                        & SoC   &    & 211    &  75.8        & 14.0       & 15                 & 99.2       \\  
    RTX 5000\footnote{\label{note1} Performance results taken from~\cite{Fang.2020}.}                    & GPU   &    & 864    &  18.5        & 61.2       & 14                 & 99.2      \\  
    Guo et al.~\cite{Guo.2019}                                                                                 & FPGA  & 32 &        &              & 0.7        &                    & 98.9       \\          
    ASIE~\cite{Kang.2020}                                                                                      & ASIC  &    &        &              & 0.001      &                    & 98.0       \\          
    SIES~\cite{Wang.2020}                                                                                      & FPGA  &    &        &              &            &                    & 99.2       \\          
    S2N2~\cite{Khodamoradi.2021}                                                                               & FPGA  &    &        &              &            &                    & 98.5       \\   
  \bottomrule
\end{tabular}
\end{center}
\footnotesize
\end{minipage}
\end{table*}

\section{Conclusions}\label{sec:Conclusions}
The SNN hardware implementation strategy developed in this work allows for low-latency
and power efficient processing of convolutional SNNs. It supports state of the art SNN
architectures thanks to the hardware implementation of max-pooling and neuronal biases.
This allows the deployed SNNs to achieve a competitive classification accuracy. 

To reduce the memory cost of the membrane potentials, a scheme for neuron multiplexing was introduced. 
In this scheme, only a small part of the SNN is simulated and only the sparse output spikes of this partial simulation are stored.
To achieve a high performance, a novel memory distribution scheme called memory
interlacing was introduced. Memory interlacing allows for a highly-parallel fine-grained distribution of memory units close to the PEs.
To exploit the high degree of sparsity, the spike events are compressed into queues. As a
result, the processing time scales with the number of occurring spikes.
Queue-based processing ensures that PEs are utilized as much as possible, even though there still is room to further improve the PE utilization. We found that in our CSNN, there were multiple channels inside the convolutional layers that never generated spikes. Thus, pruning such ``dead'' layers could lead to further improvements.
The PEs are highly pipelined to achieve a high clock frequency and to improve parallelism.
In contrast HLS-based approaches like the one of Fang et al.~\cite{Fang.2020} or S2N2~\cite{Khodamoradi.2021}, our approach is agnostic to the CSNN's architecture and can thus be implemented on an ASIC as well.
As the hardware utilization of a single convolution unit is so small, mutliple convolution units can be implemented in parallel, allowing easy scaling of throughput.
The prototype developed in this work shows a very promising performance, however, the
target SNN used here is relatively small, developed for a relatively simple benchmarking dataset. In the future we plan to implement larger SNNs and also compare our results to non-spiking implementations.

\section*{Acknowledgments}
\phantom{The paper has been partially funded by the Deutsche Forschungsgemeinschaft (DFG, German Research Foundation)~–~450987171.}

\bibliographystyle{IEEEtran}
\bibliography{bibgraph.bib}
\end{document}